\documentclass[1p,sort&compress]{elsarticle}

\usepackage{lineno,hyperref}
\modulolinenumbers[5]
\DeclareRobustCommand{\varlambda}{\text{\usefont{OML}{txmi}{m}{it}\symbol{"15}}}

\usepackage{natbib}
\usepackage{graphicx}
\usepackage{multirow}
\usepackage{relsize}
\usepackage{derivative}
\usepackage{booktabs}
\usepackage{matlab-prettifier}
\usepackage{xcolor}
\usepackage[normalem]{ulem}

\usepackage{nomencl} 
\makenomenclature
\usepackage{etoolbox} 
\renewcommand\nomgroup[1]{%
  \item[\bfseries
  \ifstrequal{#1}{A}{Abbreviations}{%
  \ifstrequal{#1}{S}{Subscripts}{%
  \ifstrequal{#1}{U}{Superscripts}}}%
]}

\usepackage{mdframed}
\usepackage{tabularx}

\makeatletter
\def\ps@pprintTitle{%
  \let\@oddhead\@empty
  \let\@evenhead\@empty
  \let\@oddfoot\@empty
  \let\@evenfoot\@oddfoot
}
\makeatother

\usepackage{amsmath,amsfonts,amssymb,amsthm,amsbsy,amsmath,bm}

\begin{document}

\begin{frontmatter}

\title{A guide to numerical dispersion curve calculations:\\ explanation, interpretation and basic \textsc{Matlab} code} 

\author[KUL,FM]{Vanessa Cool}

\author[FM,DIE]{Elke Deckers}

\author[KUL,FM]{Lucas Van Belle}

\author[KUL,FM]{Claus Claeys\corref{mycorrespondingauthor}}
\ead{claus.claeys@kuleuven.be}

\address[KUL]{KU Leuven, Department of Mechanical Engineering, Celestijnenlaan 300 - box 2420, Heverlee, Belgium}
\address[FM]{Flanders Make@KU Leuven, Belgium}
\address[DIE]{KU Leuven Campus Diepenbeek, Department of Mechanical Engineering, Wetenschapspark 27, 3590 Diepenbeek, Belgium}

\begin{abstract}
Dispersion diagrams play a crucial role in examining, analyzing and designing wave propagation in periodic structures.\ 
Despite their ubiquity and current research interest, introductory papers and reference scripting tailored to novel researchers in the field are lacking.\ 
This paper aims to address this gap, by presenting a comprehensive educational resource for researchers starting in the field of periodic structures and more specifically on the study of dispersion relations captured by dispersion surfaces or curves in dispersion diagrams.\
The objective is twofold.\ 
A first objective is to give a detailed explanation of dispersion diagrams, with graphical illustrations.\ 
Secondly, a documented \textsc{Matlab} code is provided to compute dispersion curves of 3D structures with 2D periodicity using the so-called inverse approach.\
These dispersion curves are obtained with numerical simulations using the finite element method.\
The code is written for elastic wave propagation and orthogonal periodicity directions, but can be extended to other types of linear wave propagation, non-orthogonal periodicity directions or 1D and 3D periodicity.\
The aim of this code is to serve as a starting point for novice researchers in the field, to facilitate their understanding of different aspects of dispersion diagrams and serve as a stepping stone in their future research.
\end{abstract}


\end{frontmatter}

\begin{mdframed}
\section*{Abbreviations}
\begin{flushleft}
  \renewcommand{\arraystretch}{1.1}
  \begin{tabularx}{\textwidth}{@{}p{20mm}X@{}}
        BMS     &   Bloch mode synthesis\\
        DOF     &   Degree-of-freedom\\
        FBZ     &   First Brillouin zone \\
        FE      &   Finite elements\\
        GBMS    &   Generalized Bloch mode synthesis\\
        IBC     &   Irreducible Brillouin contour\\
        IBZ     &   Irreducible Brillouin zone \\
         MOR     &   Model order reduction\\
        PCG     &   Plane crystallographic group \\
        STL     &   Sound transmission loss\\
        UC      &   Unit cell\\
        WFE     &   Wave finite elements\\
        & \\
  \end{tabularx}
\end{flushleft}
\end{mdframed}

\section{Introduction}
This paper is intended for educational use and for researchers who are new to the field of wave propagation in periodic media, particularly when dealing with dispersion curves in dispersion diagrams.\ 
The objective of this paper is twofold: to facilitate a more accessible approach to dispersion curve calculations and to offer guidance for students and novice researchers in comprehending and cultivating an intuition for interpreting the dispersion diagrams.\
To accomplish this dual objective, the paper is organized into two parts: (i)~a theoretical and modeling part, supported by ample visualizations, intended to equip the reader with crucial insights into the interpretation of dispersion diagrams - a perspective that is presently scarce in existing literature; (ii)~a documented \textsc{Matlab} code, providing a tool that can directly be used to explore various case studies and which can be straightforwardly extended.\
While the discussed concepts are generally applicable, elastodynamic wave propagation in 2D periodic media is considered as a use-case throughout the manuscript.\
Summarized, the focus of this paper is to discuss and provide the theoretical background behind dispersion diagrams and the implementation to derive the dispersion curves in an accessible manner, while it is not the purpose to provide a review of all current literature on the topic.\ 

Dispersion diagrams serve as a frequently employed tool in the examination of wave propagation phenomena.\ 
They are visual representations of dispersion relations, which give the relation between frequency and wave number for various waves propagating through the system.\ 
Their popularity has grown significantly, especially in the investigation of intricate wave phenomena.\ 
In particular in the context of periodic media, the use of dispersion diagrams, or band diagrams, finds its origin in solid-state physics, where band structures describe
the existence of energy bands in which electrons can reside, separated by forbidden bandgaps.\ 
The seminal work of Bloch \cite{bloch1929quantenmechanik}, applying Floquet \cite{floquet1883equations} theory to periodic crystal lattices, provides a powerful theory for band structure computation of periodic media.

Of course, curiosity has driven researchers to search for bandgaps in other branches of physics in order to manipulate wave propagation.\ 
Photonic crystals were for instance proposed, which are periodic structures with bandgaps for electromagnetic wave propagation \cite{Ho1990, Yablonovitch1991}.\ 
In this case, bandgaps or stop bands are frequency ranges in which freely propagating waves are forbidden \cite{brillouin1946wave}.\ 
The analogies between the electromagnetic, acoustic and elastic wave equations have meanwhile led to the emergence of phononic crystals and metamaterials with stop bands for acoustic and for elastic wave propagation.

Hence, for this wide variety of periodic structures, dispersion diagrams or band diagrams provide a useful tool to identify and gain insight in stop bands, but also in other phenomena such as wave veering, locking and coupling \cite{vasileiadis2021progress,zhou2012elastic,liu2011metamaterials,lopez2003materials,hussein2014dynamics} with, among others, applications in acoustic lenses and wave propagation control~\cite{zhu2017two,lemoult2013wave}.\ 
Furthermore, as the interest in manipulating and precisely engineering wave phenomena grows, along with the expanding possibilities to do so through concepts like phononic crystals and metamaterials, dispersion diagrams have become a widespread and go-to tool in the analysis and design of these (often) periodic, complex media.

While wave propagation in very simple systems can be described by analytical dispersion relations, this is no longer the case for the nowadays considered plenitude of  complex periodic structures.\ 
Instead, a variety of numerical approaches for dispersion relation computations have emerged, including plane wave expansion, multiple scattering, finite difference and finite element (FE) methods.\ 
Although the computation of dispersion curves or surfaces representing the dispersion relation has been discussed to some extent in literature, e.g.\ \cite{mace_2008_techn}, currently there is no adequate educational paper that focuses on introducing novice researchers to the interpretation, reading, and numerical generation of the dispersion curves in dispersion diagrams.\ 
This work aims to close this gap.\
Given the background of the authors' metamaterial research, the paper approaches the explanation of dispersion diagrams in the context of elastodynamics using FE modeling.\ 
Nevertheless, the introduced concepts are also more generally applicable to other fields of research.\

The remainder of this paper is structured as follows.\
 Sec.~\ref{sec:background} provides an introduction into dispersion relations and wave propagation tailored to novel researchers in the field.\
Sec.~\ref{sec:modeling} introduces the modeling of infinite periodic structures.\
Next, Sec.~\ref{sec:disp_curves} explains how to go from the dispersion eigenvalue problem to the commonly used dispersion diagrams.\
Afterwards, a basic \textsc{Matlab} code to compute dispersion curves is discussed in Sec.~\ref{sec:Matlab_impl_new}  together with a discussion of the dispersion diagrams for three use-cases.\ 
The complete \textsc{Matlab} code is given in \ref{app:matlab}, while a detailed discussion on the implementation is given in \ref{sec:matlab_impl} and inspiration for possible extensions is provided in \ref{sec:ext}.\
The main contributions and key points discussed in this paper are summarized in Sec.~\ref{sec:concl}.\

\section{Introduction into dispersion relations and wave propagation}\label{sec:background}
Before discussing dispersion diagrams of 2D periodic media, this section provides a basic introduction to elastic wave propagation, the associated key fundamental concepts and the very basics on dispersion relations, surfaces and curves.\ 
As it is not the intention to provide a rigorous and elaborate overview, but rather to enable a better understanding of the rest of the manuscript, the explanation in this section will be limited to the illustrative example of wave propagation in thin plates.\
The reader who is not familiar with the general concepts of wave propagation, in particular wave propagation in plates and elastic media, is referred to reference literature for a more in-depth explanation~\cite{fahy2007sound,cremer1974structure,hambric2006structural}.

\begin{figure*}
	\centering
	\includegraphics[width = \linewidth]{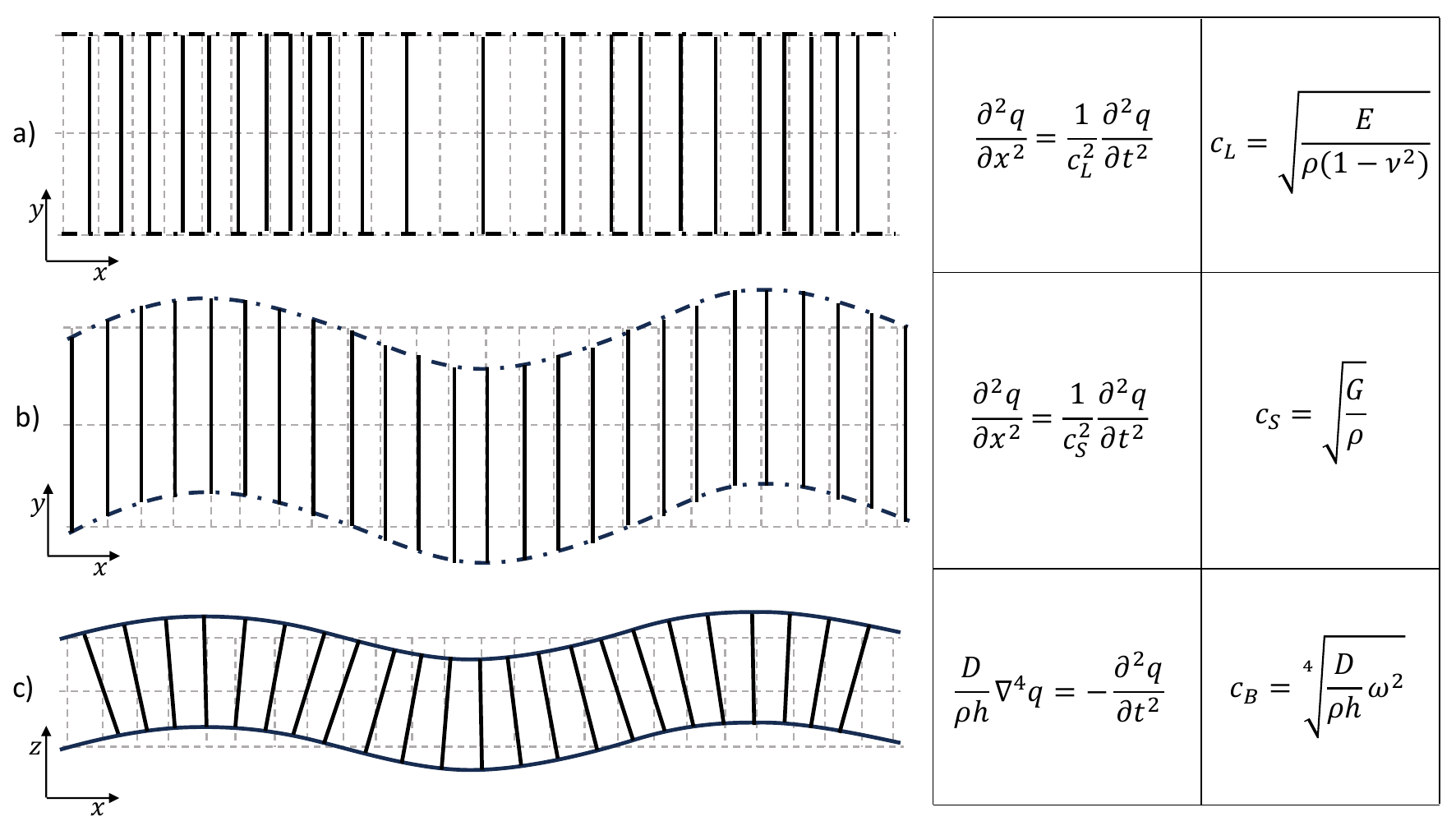}
	\caption{Visualization of the different wave types, considering propagation along the $x$-direction in the $xy$- plane (adapted from~\cite{fahy2007sound}): a)~top view of the plate with a longitudinal wave visualized, b)~top view of the plate with a shear wave visualized and c)~cross-section of the plate with a bending wave visualized.\ The dotted gray lines represent the undeformed state.\ The corresponding governing wave equations and expressions for the wave speed are given on the right in which $c_L$, $c_S$ and $c_B$ are the wave speeds for the longitudinal, shear and bending wave, respectively, while E, G and D represent the elastic Young's modulus, the elastic shear modulus and the flexural rigidity.}
	\label{fig:wave_types}
\end{figure*}

When a disturbance of a mechanical quantity, i.e.~an elastic deformation, is transmitted to surrounding particles in the medium, thereby transferring energy throughout the medium, mechanical waves are said to travel through that medium.\
In the case of an isotropic, homogeneous, thin, flat plate with constant thickness, three different types of elastic waves can travel along its in-plane directions: longitudinal, shear and bending waves (Fig.~\ref{fig:wave_types}).\ 
This categorization relates to the shape and orientation of the deformation pattern which propagates along the wave's travel direction: i)~longitudinal waves, also called compression waves, are characterized by a particle motion parallel to the direction of the wave propagation, compressing and expanding the plate in that direction (Fig.\ \ref{fig:wave_types}a).\ 
This is similar to acoustic waves in fluids.\ 
ii)~Shear waves can also propagate through the plate, since solids can resist shear deformation, contrary to fluids.\ 
In this case, the particles undergo an in-plane sliding motion, in the direction perpendicular to the wave propagation direction (Fig.\ \ref{fig:wave_types}b).\ 
iii)~Bending waves, also called flexural waves, differ from the other two wave types in that they are characterized by transverse out-of-plane deformation along the wave propagation direction (Fig.\ \ref{fig:wave_types}c).\
In this example, the assumption is made that the plate is of infinite extent in its in-plane directions.\ 
This idealization allows to study waves without considering interactions with boundaries, and hence, simplifies the mathematical models and their analyses.\ 

The wave equations and wave speeds for these three wave types in thin plates are given in Fig.~\ref{fig:wave_types}.\ 
Contrary to the longitudinal and shear waves, the bending wave speed depends on the frequency: these waves are said to be \textit{dispersive}.\ 
Nevertheless, the wave equations in Fig.~\ref{fig:wave_types} are linear, resulting in natural solutions which can be represented by harmonic waves of the following form:
\begin{equation}
\label{eq:harm_wave}
    q = A e^{\mathrm{i}\omega t} e^{-\mathrm{i} (k_x x + k_y y) },
\end{equation}
with $A$ the wave amplitude, $\omega$ the angular frequency, $t$ the time, and $k_x$, $k_y$ the wavenumbers in the $x$- and $y$-direction, respectively.\
On the one hand, the frequency $f$, obtained from $\omega = 2\pi f$, defines the temporal characteristic of the wave and gives a measure of the number of oscillations of a wave per unit of time, i.e.~$ \omega = 2\pi/T $ with $T$ the period of the wave.\
On the other hand, the wavenumbers $k_x$, $k_y$ characterize the spatial behavior of a wave in the $x$- and $y$-direction, giving a measure of the number of oscillations of a wave per unit of distance, i.e.~$ k_x = 2\pi/\varlambda_x$ and $ k_y = 2\pi/\varlambda_y$ with $\varlambda_x$ and $\varlambda_y$ the wavelength in the $x$- and $y$-direction, respectively.\ 
Together, these wavenumbers constitute the wave vector $\textbf{k}= (k_x,k_y)$. With the wavelength representing the \textit{spatial period}, the wavenumber can thus be interpreted as the \textit{spatial frequency}.

\begin{figure*}
	\centering
	\includegraphics[width = \linewidth]{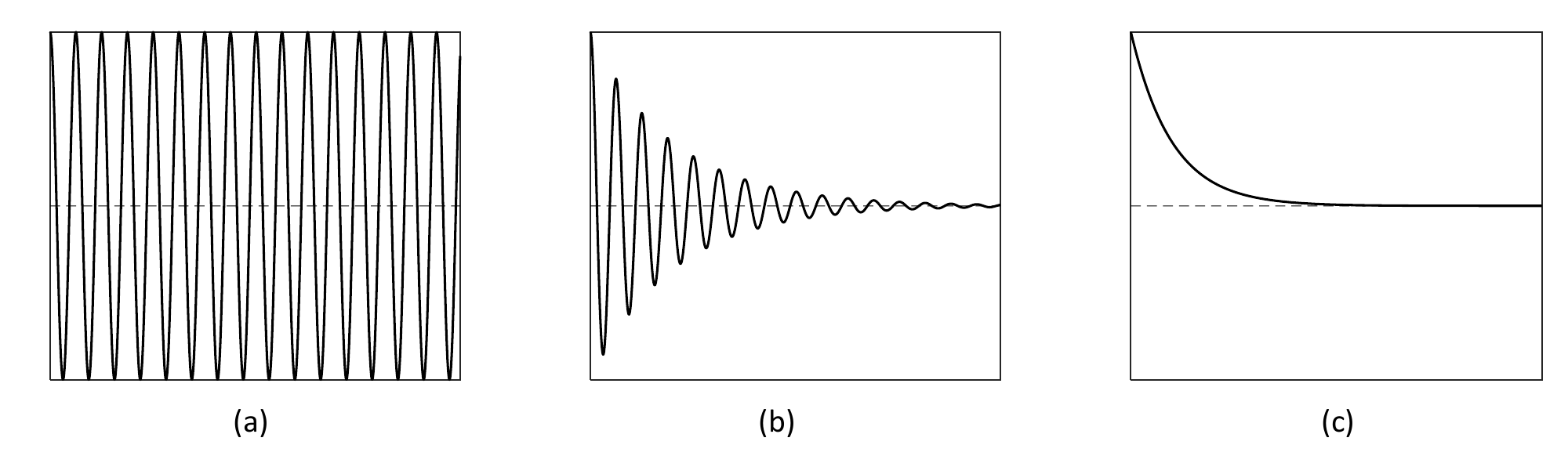}
	\caption{ Visualization of a a)~propagating, b)~decaying and c)~evanescent wave.\ 
 In the case of temporal waves, the horizontal axis represents time and the vertical axis represents the amplitude of one element of the plate.\
 In the case of spatial waves, the horizontal axis represents space, i.e., the position of the different elements of the plates, and the vertical axis represents the amplitude of each element.\ 
 It should be noted that in this view, abstraction is made of the relative motion across a cross-section of the plate.} 
	\label{fig:wave_types2}
\end{figure*}

Dispersion diagrams are visual representations of the dispersion relation, i.e.~the relation between frequency and wave number.\ 
Hence, for each wave type, one needs an expression relating $\omega$ and $k$ before being able to visualize it.\ 
The angular frequency $\omega$, wavenumber $k$ and wave speed $c$ (also named \textit{phase velocity}) of a wave are related as $k=\omega/c$ ~\cite{fahy2007sound}.\ 
Hence, based on the expressions of the wave speed in Fig.~\ref{fig:wave_types} for the three wave types, this gives rise to the following \textit{dispersion relations}:
\begin{equation}
\label{eq:analyt_kw}
    k_L^2 = \frac{\omega^2\rho (1-\nu^2)}{E},\quad k_S^2 = \frac{2\omega^2\rho (1+\nu)}{E},\quad k_B^4 = \frac{12\omega^2\rho(1-\nu^2)}{Eh^2},
\end{equation}
with $k_L$, $k_S$ and $k_B$ the wavenumbers for the longitudinal, shear and bending wave types along a certain wave travel direction.\ 
In case one aims to visualize these dispersion relations, either $\omega$ or $k$ can be chosen as the dependent variable.\ 
Although the main interest often lies in dispersion relations for time-harmonic freely propagating waves, corresponding to real $\omega$ and $k$, in principle either can take real, complex or imaginary values, in particular in presence of damping, which gives rise to, respectively 
\textit{propagating}, \textit{decaying} and \textit{evanescent} waves (Fig.~\ref{fig:wave_types2}).\
Propagating waves travel through the medium without loss of energy.\ 
Decaying waves are characterized by a decreasing amplitude due to energy dissipation or attenuation, while evanescent waves cannot transmit energy through the medium.\
The latter two wave types decay, respectively, in the space or time domain whenever the frequency or wavenumber becomes complex, i.e.~the $x$-axis of Fig.~\ref{fig:wave_types2} represents, respectively, space or time.\
The nature of these different waves can also be linked to Eq.~(\ref{eq:harm_wave}).\

\begin{figure*}
	\centering
	\includegraphics[width = \linewidth]{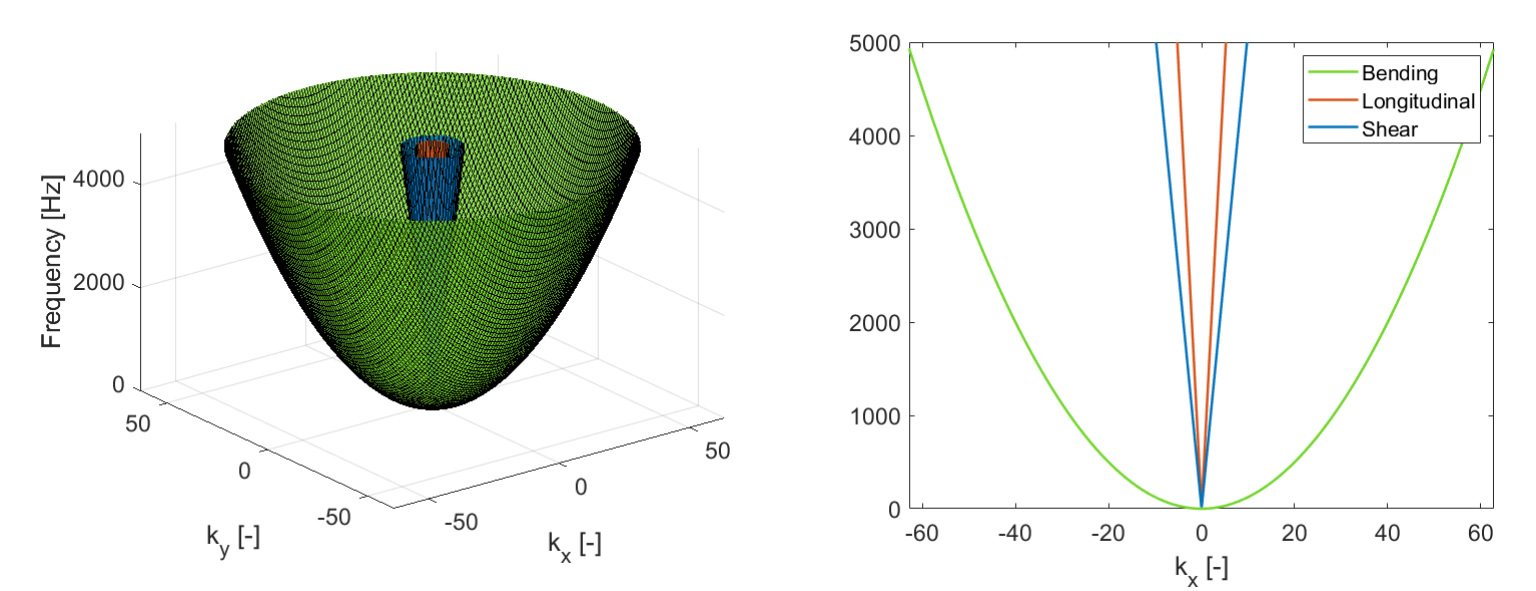}
	\caption{ Visualized time-harmonic freely propagating bending (green), longitudinal (red) and shear (blue) wave solutions in a thin plate: dispersion surfaces (left), dispersion curves of wave propagation in the $x$-direction (right). }
	\label{fig:an_DC}
\end{figure*}

Considering now again the dispersion relations of Eq.~(\ref{eq:analyt_kw}) for an isotropic homogeneous thin plate, it is clear that these relations hold for any propagation direction in the $xy$-plane, with $k^2 = k_x^2+k_y^2$.\ 
Reformulating these dispersion relations in terms of $\omega$ as function of $k$ and focusing on the time-harmonic freely propagating wave solutions of these dispersion relations, the corresponding real $\omega$ and $k$ solutions can be plotted in $(k_x,k_y,\omega)$ space.\
Fig.~\ref{fig:an_DC} shows a plot of the dispersion relation, assuming that the thin homogeneous, isotropic plate under study has a thickness $h=0.005$~m and consists of steel with material properties: Young's modulus $E=210$~GPa, density $\rho=7800$~kg/m$^3$ and Poisson's ratio $\nu=0.3$.\ 
The longitudinal and shear wave solutions correspond to conical surfaces originating at zero, and the \textit{dispersive} bending wave solution corresponds to a paraboloid originating at zero.\ 
These surfaces are called \textit{dispersion surfaces}.\ 
If now time-harmonic freely propagating wave solutions along one specific propagation direction would be of interest, this would correspond to taking a slice of the dispersion surfaces along this direction.\ 
This is shown in Fig.~\ref{fig:an_DC} for wave propagation in the $x$-direction, taking a slice at $k_y=0$.\ 
The corresponding curves are generally denoted as the \textit{dispersion curves} and are visualized in a \textit{dispersion diagram}. 

Dispersion surfaces, and dispersion curves, clearly provide a convenient representation of the wave propagation in infinite structures.\ 
Each point represents a wave which travels through the structure at a certain frequency, with a certain wave vector.\ 
Although real-life structures are of course finite, these dispersion surfaces and curves for infinite structures provide an excellent and insightful basis for understanding and interpreting wave behavior in finite structure counterparts.

For simple homogeneous structures, the dispersion relations can be derived analytically.\ 
However, in recent years, an increasing interest has grown in analyzing and designing complex periodic media which can enable tailoring the wave propagation.\ 
To facilitate the analysis of these periodic media, infinite periodicity is typically assumed such that dispersion curves can again be computed to predict the wave propagation in the infinite periodic structure as the basis for understanding the behavior of their finite periodic counterparts.\ 
However, due to the often high complexity of the underlying structures, analytical derivations of dispersion relations rapidly become cumbersome, if not impossible, to derive and numerical methods are often resorted to instead.\
Therefore, the following section discusses the modeling of infinite periodic structures and how to obtain the dispersion curves for these structures. \\

\section{Modeling infinite periodic structures}
\label{sec:modeling}
Infinite periodic structures can be fully described by a reference unit cell (UC), which is the smallest non-repetitive part.\
Based on a model of this UC, the dispersion curves for an infinite periodic structure can be calculated.\
In this section, a brief introduction to Bloch's theorem{\footnote{Note that in the literature the terms Bloch's theorem, Bloch-Floquet theorem or Floquet-Bloch principle are all applied.\ In this work, the term Bloch's theorem is used.}} is given, which is the fundamental theorem that allows studying wave propagation of infinite (periodic) structures based on their UC.\ 
This theorem is afterwards further detailed for the case of FE-based UC modeling.\

\begin{figure*}
	\centering
	\includegraphics[width = \linewidth]{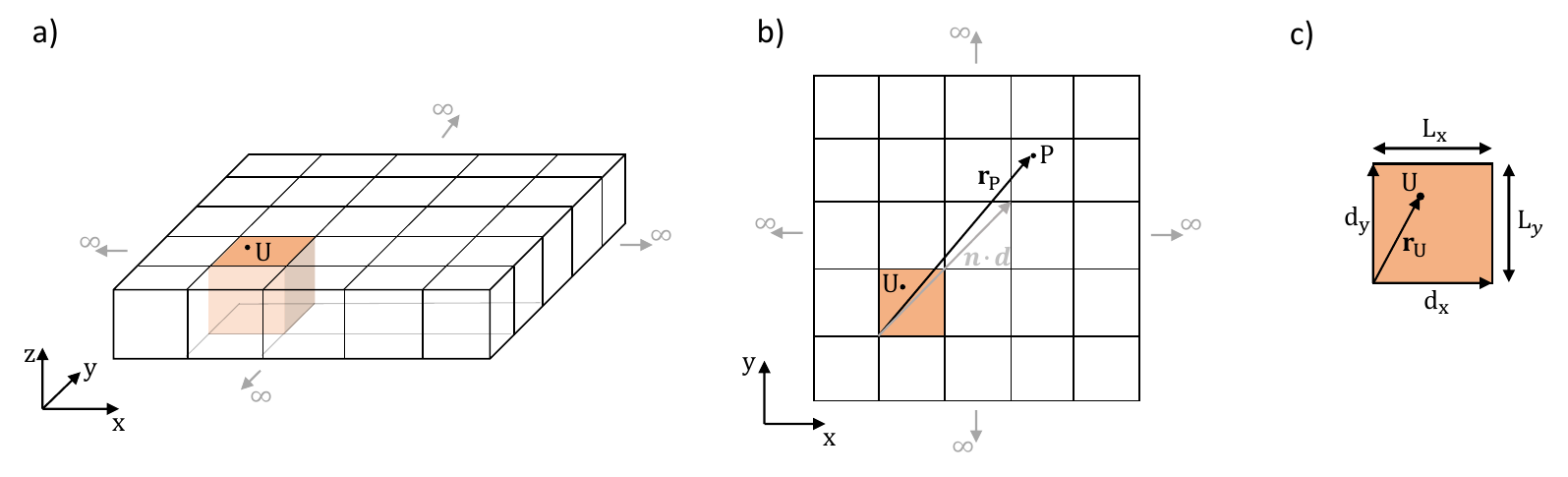}
	\caption{Schematic representation of an infinite 2D periodic structure. a)~3D representation, one UC is highlighted in orange.\ b)~Top view of the infinite periodic structure.\ c)~Corresponding UC, as the fundamental building block of the infinite periodic structure.}
	\label{fig:theory_BF}
\end{figure*}

\newpage
\subsection{Bloch's theorem}
\label{subsec:BF}
This paper focuses specifically on rectangular structures with 2D periodicity as schematically represented in Fig.~\ref{fig:theory_BF}.\
The structures are represented by a UC with dimensions $L_x \times L_y$, indicated in orange in Fig.~\ref{fig:theory_BF}, which periodically repeats in the $xy$-plane along two orthogonal directions, denoted by $\mathbf{d}_x$ and $\mathbf{d}_y$.\
Since the structure is periodic, the position of each point $\mathrm{P}$ in the structure, denoted as $\mathbf{r}_\mathrm{P}$, can be expressed with respect to the corresponding point $\mathrm{U}$ in the reference UC, translated $n_x$ cells along $\mathbf{d}_x$ and $n_y$ cells along $\mathbf{d}_y$:
\begin{equation}
\label{eq:rP}
    \mathbf{r}_\mathrm{P} = \mathbf{r}_\mathrm{U} + n_x \mathbf{d}_x + n_y \mathbf{d}_y 
     = \mathbf{r}_\mathrm{U} + \mathbf{n} \cdot \mathbf{d}, 
\end{equation}
in which $\mathbf{n} = (n_x,n_y)$, $\mathbf{d}=(\mathbf{d}_x,\mathbf{d}_y)$ and '$\cdot$' represents the scalar product.

Bloch's theorem~\cite{bloch1929quantenmechanik,brillouin1946wave} governs the wave propagation in periodic media, stating the solution can be expressed as plane waves modulated with periodic functions.\ As a consequence of this theorem, wave propagation in an infinite periodic structure can be expressed in terms of the response of a reference UC, and an exponential term defining the relative amplitude and phase change as the wave propagates from one UC to the next.\
The change in phase and wave amplitude occurring from UC to UC does not depend on the UC location within the periodic structure.\
Hence, the wave response at a point $\mathrm{P}$ of the periodic structure, denoted as $\mathbf{q}(\mathbf{r}_\mathrm{P},\mathbf{k},\omega)$\, can be expressed as function of the response at the corresponding point $\mathrm{U}$ in the reference UC, denoted as $\mathbf{q}_{ref}(\mathbf{r}_\mathrm{U},\mathbf{k},\omega)$:
\begin{equation}
\label{eq:qk}
    \mathbf{q}(\mathbf{r}_\mathrm{P},\mathbf{k},\omega) =  \mathbf{q}_{ref}(\mathbf{r}_\mathrm{U},\mathbf{k},\omega) e^{\mathrm{i}\mathbf{k} \cdot (n_x \mathbf{d}_x + n_y \mathbf{d}_y)}.
\end{equation}
As introduced in Sec.~\ref{sec:background}, $\mathbf{k} = (k_x,k_y)$ is the (complex) wave vector, composed of the wave numbers in the  $x$- and $y$-direction.\ 
The real and imaginary part represent, respectively, the amplitude decay and phase change of a wave per meter in each direction.\
In~\ref{app:Bloch}, more details are given on how Eq.~(\ref{eq:qk}) is obtained with Bloch's theorem.\
Eq.~(\ref{eq:qk}) is often written in terms of a vector $\bm{\mu}$ which consists of the (frequency dependent) wave propagation constants $(\mu_x, \mu_y)$:
\begin{equation}
\label{eq:qmu}
    \mathbf{q}(\mathbf{r}_\mathrm{P},\boldsymbol\mu ,\omega) =  \mathbf{q}_{ref}(\mathbf{r}_\mathrm{U}, \boldsymbol\mu ,\omega) e^{\mathrm{i} \mathbf{\boldsymbol\mu} \cdot \mathbf{n}}, 
\end{equation}
with 
\begin{equation}
\label{eq:mu}
    \boldsymbol\mu = (\mu_x, \mu_y) = (\mathbf{k} \cdot \mathbf{d}_x, \mathbf{k} \cdot \mathbf{d}_y) = (k_x L_x, k_y L_y).
\end{equation}
The introduction of these propagation constants brings the advantage that wave propagation is expressed in terms of variation per UC instead of per meter  as was the case when using the wave vector $\textbf{k}$.\ 
The real and imaginary part of the (complex) propagation constants now denote, respectively, the relative change in amplitude and the change in phase when moving across one UC in the $x$- or $y$-direction.\

\subsection{FE UC modeling}
\begin{figure*}
	\centering
	\includegraphics[width = \linewidth]{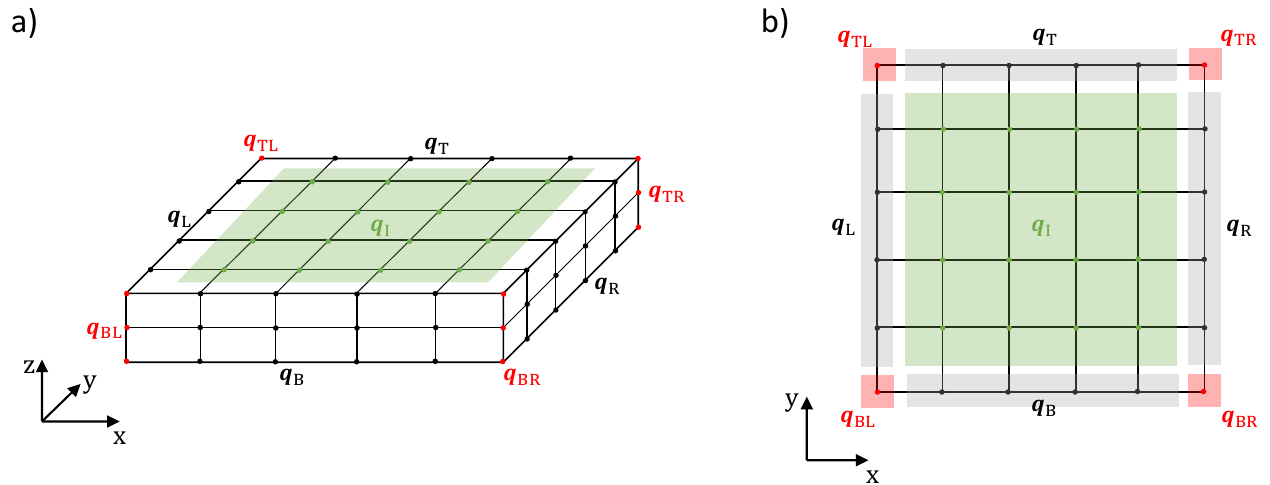}
	\caption{Visualization of the different node groups. a)~3D UC representation discretized with $5\times 5 \times 2$ elements and notation of the different node groups. b)~Corresponding top view.}
	\label{fig:theory_nodes}
\end{figure*}

Considering the wave motion in the infinite periodic structure, Bloch's theorem states that the displacements and forces on the boundaries of a UC are scaled with a factor $e^{\mathrm{i} \mu_x}$ and $e^{\mathrm{i} \mu_y}$ when moving from one UC to the next in the $\mathbf{d}_x$-, $\mathbf{d}_y$-direction, respectively.\
Hence, to model the wave propagation based on a (possibly geometrically complex) UC, a suitable UC modeling strategy is required.\
Several methods exist to model the UC, i.e.~the plane wave expansion, multiple scattering, finite difference and FE methods.\
Due to its versatility to discretize complex geometries, the FE method is a popular and commonly used method~\cite{mead1996wave}.\
Using the standard FE method{\footnote{Although adapted FE methods exist, which start from a modified weak formulation embedding Bloch's theorem, the implementation of this method is less straightforward and thus not considered here.}}, Bloch's theorem can be translated into appropriate periodic boundary conditions applied on the UC.\
Fig.~\ref{fig:theory_nodes}a shows a UC discretized with $5 \times 5 \times 2$ solid elements.\
The resulting discrete linear system of equations of the UC with $N$~degrees-of-freedom (DOFs), assuming time-harmonic motion with $e^{\mathrm{i}\omega t}$ dependency, is given here directly in the frequency domain without its time-domain counterpart~\cite{mead1973general,mead1996wave}:
\begin{equation}
\label{eq:eq_motion}
    (\mathbf{K}-\omega^2 \mathbf{M}) \mathbf{q}=\mathbf{f},
\end{equation}
with $\mathbf{K}$, $\mathbf{M}$ $\in \mathbb{R}^{N \times N}$ the stiffness and mass system matrices, and $\mathbf{q}$, $\mathbf{f}$ $\in \mathbb{R}^{N \times 1}$ the nodal displacement DOFs of the UC and the external nodal forces applied on the UC, respectively.\
Note that no structural or viscous damping is included here, although including damping is possible \cite{van2017impact}.\
To simplify the application of the periodicity boundary conditions, the UC DOFs $\mathbf{q}$ are partitioned according to the UC boundaries and interior (Fig.~\ref{fig:theory_nodes}):
\begin{equation}
\label{eq:BF_BCs}
\mathbf{q} =
\left[ 
\mathbf{q}_I^T  \hspace{0.08cm} 
\mathbf{q}_L^T  \hspace{0.08cm}
\mathbf{q}_R^T  \hspace{0.08cm}
\mathbf{q}_B^T  \hspace{0.08cm}
\mathbf{q}_T^T  \hspace{0.08cm}
\mathbf{q}_{BL}^T \hspace{0.08cm}
\mathbf{q}_{BR}^T \hspace{0.08cm}
\mathbf{q}_{TL}^T \hspace{0.08cm}
\mathbf{q}_{TR}^T  \hspace{0.08cm}
\right]^T,
\end{equation}
in which subscripts $I,L,R,B,T$ denote the interior, left, right, bottom and top parts, respectively, while $BL,BR, TL$ and $TR$ are the corner DOFs.\
By defining{\footnote{Note that $\lambda_x$ and $\lambda_y$ are sometimes denoted as the propagation constants, e.g.~\cite{mace2008modelling}, this is however not done in this paper to avoid confusion with $\mu_x$ and $\mu_y$.}} $\lambda_x = e^{\mathrm{i}\mu_x}$ and $\lambda_y = e^{\mathrm{i}\mu_y}$, Bloch's theorem results in the following relations for the UC DOFs~\cite{langley1993note}:
\begin{equation}
\begin{gathered}
    \mathbf{q}_R = \lambda_x \mathbf{q}_L,
    \hspace{0.3cm}
    \mathbf{q}_T = \lambda_y \mathbf{q}_B, \\
    \mathbf{q}_{BR} = \lambda_x \mathbf{q}_{BL}, 
    \hspace{0.3cm}
    \mathbf{q}_{TL} = \lambda_y \mathbf{q}_{BL}, 
    \hspace{0.3cm}
    \mathbf{q}_{TR} = \lambda_x \lambda_y \mathbf{q}_{BL}. 
    \hspace{0.3cm}    
\end{gathered}
\end{equation}
In matrix-vector notation this reads as~follows \cite{mace2008modelling}:
\begin{equation}
\label{eq:BF_xy}
    \mathbf{q} = 
    \left[ \begin{matrix} 
\mathbf{I}  & \mathbf{0}    & \mathbf{0}     & \mathbf{0}  \\ 
\mathbf{0} & \mathbf{I} & \mathbf{0}       & \mathbf{0}  \\
\mathbf{0} & \lambda_x \mathbf{I} & \mathbf{0}       & \mathbf{0}  \\
\mathbf{0} & \mathbf{0}    & \mathbf{I}    & \mathbf{0}  \\
\mathbf{0} & \mathbf{0}    & \lambda_y \mathbf{I}    & \mathbf{0}  \\
\mathbf{0} & \mathbf{0}    & \mathbf{0}    & \mathbf{I}  \\
\mathbf{0} & \mathbf{0}    & \mathbf{0}    & \lambda_x \mathbf{I}  \\
\mathbf{0} & \mathbf{0}    & \mathbf{0}    & \lambda_y \mathbf{I}  \\
\mathbf{0} & \mathbf{0}    & \mathbf{0}    & \lambda_x \lambda_y \mathbf{I}  \\
\end{matrix} \right]
    \left[ \begin{matrix}
    \mathbf{q}_{I} \\
    \mathbf{q}_L  \\
    \mathbf{q}_{B} \\
    \mathbf{q}_{BL}  \\
    \end{matrix} \right] = \mathbf{R}
    \tilde{\mathbf{q}},
\end{equation}
in which $\mathbf{R}$ is the periodicity matrix and $\tilde{\mathbf{q}}$ is the periodic DOF vector.\
\begin{figure*}
	\centering
	\includegraphics[width = 0.5\linewidth]{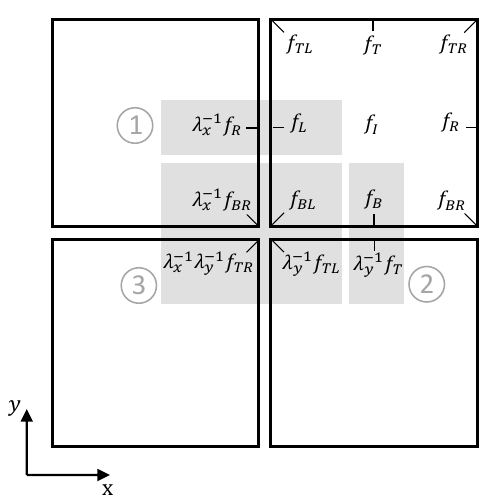}
	\caption{Equilibrium of the generalized forces at \textcircled{1} the left boundary, \textcircled{2} the bottom boundary and \textcircled{3} the bottom-left corner of consecutive UCs~\cite{langley1993note}.}
	\label{fig:force_balance}
\end{figure*}
An analogous relation can be written to express the force balance at the left boundary, bottom boundary and bottom-left corner of the UC~\cite{langley1993note} (Fig.~\ref{fig:force_balance}):
\begin{equation}
\label{eq:force_bal}
\begin{aligned}
&\textcircled{1} \hspace{0.7cm} \mathbf{0}= \mathbf{f}_L + \lambda_x^{-1}\mathbf{f}_R, \\
&\textcircled{2} \hspace{0.7cm} \mathbf{0}= \mathbf{f}_B + \lambda_y^{-1}\mathbf{f}_T, \\
&\textcircled{3} \hspace{0.7cm} \mathbf{0}= \mathbf{f}_{BL} + \lambda_x^{-1}\mathbf{f}_{BR} + \lambda_y^{-1}\mathbf{f}_{TL} + \lambda_x^{-1}\lambda_y^{-1}\mathbf{f}_{TR}.\\
\end{aligned}
\end{equation}
More details on the derivation of these equations and the presence of the inverse of $\lambda_x$ and $\lambda_y$ are given in \ref{app:force}.\
Eqs.~(\ref{eq:force_bal}) can also be rewritten in matrix-vector notation as follows:
\begin{equation}
\label{eq:BF_force}
    \left[ \begin{matrix} 
\mathbf{I}  & \mathbf{0} & \mathbf{0} & \mathbf{0} & \mathbf{0} & \mathbf{0} & \mathbf{0} & \mathbf{0} & \mathbf{0} \\
\mathbf{0} & \mathbf{I} & \lambda_x^{-1} \mathbf{I} & \mathbf{0} & \mathbf{0} & \mathbf{0} & \mathbf{0} & \mathbf{0} & \mathbf{0} \\
\mathbf{0} & \mathbf{0} & \mathbf{0} & \mathbf{I} & \lambda_y^{-1} \mathbf{I} &  \mathbf{0} & \mathbf{0} & \mathbf{0} & \mathbf{0} \\
\mathbf{0} & \mathbf{0} & \mathbf{0} & \mathbf{0} & \mathbf{0} & \mathbf{I} &  \lambda_x^{-1} \mathbf{I} &  \lambda_y^{-1} \mathbf{I}  & \lambda_x^{-1} \lambda_y^{-1} \mathbf{I}  \\
\end{matrix} \right] 
\left[ \begin{matrix}
    \mathbf{f}_{I} \\
    \mathbf{f}_L  \\
    \mathbf{f}_R  \\
    \mathbf{f}_{B} \\
    \mathbf{f}_T  \\
    \mathbf{f}_{BL}  \\
    \mathbf{f}_{BR}  \\
    \mathbf{f}_{TL}  \\
    \mathbf{f}_{TR}  \\
    \end{matrix} \right]= \mathbf{R}' \mathbf{f} = 
    \left[ \begin{matrix}
    \mathbf{f}_{I} \\
    \mathbf{0}  \\
    \mathbf{0}  \\
    \mathbf{0}  \\
    \end{matrix} \right].
\end{equation}
Note that the periodic force matrix is denoted by $\mathbf{R}'$.\ 
Whenever real ($\mu_x,\mu_y$) are used, $\mathbf{R}'$ equals the conjugate transpose of $\mathbf{R}$.\
Substituting Eq.~(\ref{eq:BF_xy}) into Eq.~(\ref{eq:eq_motion}), pre-multiplying with $\mathbf{R}'$ and assuming that no external forces are applied to the internal DOFs ($\mathbf{f}_I = \mathbf{0}$), the following generalized dispersion eigenvalue problem is obtained~\cite{mace2008modelling,hussein2014dynamics}:
\begin{equation}
\label{eq:EVP_BF}
    (\mathbf{\Tilde{K}}-\omega^2 \mathbf{\Tilde{M}})\mathbf{\Tilde{q}} = \mathbf{0}, 
\end{equation}
in which $\mathbf{\Tilde{K}},\mathbf{\Tilde{M}}$ are a function of ($\mu_x,\mu_y$) and given by:
\begin{equation}
\label{eq:Proj}
    \mathbf{\Tilde{K}} = \mathbf{R}' \mathbf{K}\mathbf{R},  \hspace{1cm} 
    \mathbf{\Tilde{M}} = \mathbf{R}' \mathbf{M}\mathbf{R}. 
\end{equation}
By solving this dispersion eigenvalue problem, the dispersion curves can be obtained, as will be further explained in the following section.\
Note that the corresponding eigenvalue problem contains three unknowns, namely $\omega$, $\mu_x$ and $\mu_y$, which can all be complex.\
As a consequence, different approaches can be followed to solve the eigenvalue problem~\cite{mace2008modelling,mace2014discussion}.\
The two most commonly applied solution approaches are:
\begin{itemize}
    \item \textbf{Inverse approach} $\omega(\boldsymbol\mu)$ - real wave propagation constants ($\mu_x, \mu_y$) are imposed and the eigenvalue problem is solved toward the frequencies.\ 
    The method thus considers freely propagating waves.\ 
    The resulting frequencies can generally be real, imaginary or complex representing freely propagating, (temporally) evanescent or (temporally) decaying waves~\cite{manconi2010estimation}, respectively, cf.~Fig.~\ref{fig:wave_types2} with the horizontal axis representing time.\ 
    Since the inverse approach is typically used for undamped UCs, real frequencies will result when imposing real propagation constants.\
 Note that in this manner, stop bands or bandgaps can easily be identified as they correspond to frequency ranges without freely propagating waves.\ 
    \item \textbf{Direct approach} $\boldsymbol\mu(\omega)$ - real $\omega$ are imposed and the eigenvalue problem is solved to the wave propagation constants.\
    Therefore the assumption of time-harmonic wave propagation is made.\ 
    The resulting wave propagation constants can be real, imaginary or complex representing freely propagating, spatially evanescent and spatially decaying wave propagation, respectively, cf.~Fig.~\ref{fig:wave_types2} with the horizontal axis representing space.\
    While the implementation of the direct approach is straightforward for 1D periodic problems~\cite{mace2005finite}, it becomes cumbersome for 2D periodic problems since two unknowns remain after imposing $\omega$.\ Different approaches are described in the literature to resolve this~\cite{mace2008modelling,meng2018new}.\
\end{itemize}
Next to the inverse and direct approach, also mixed approaches exist~\cite{mace2014discussion} or approaches which simultaneously solve for complex $\omega$ and $\boldsymbol\mu$~\cite{frazier2016generalized}.\ 
In this work, the inverse approach is further applied to obtain the dispersion curves of freely propagating waves for undamped structures.\
Understanding these dispersion curves allows to gain insight which can be further built upon when investigating damped structures, by subsequently making use of the direct approach.\

\begin{figure*}[h!]
	\centering
	\includegraphics[width = \linewidth]{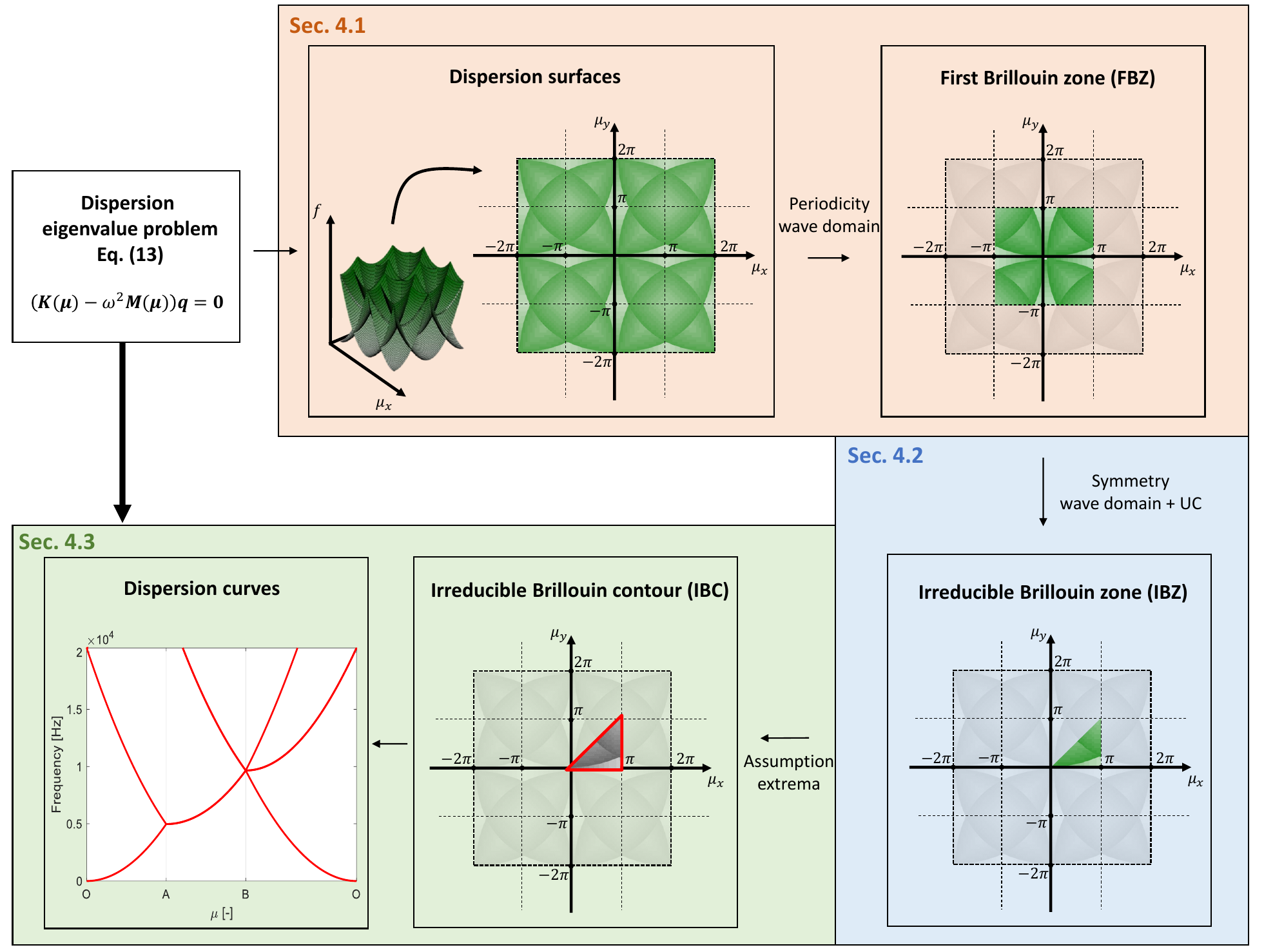}
	\caption{Overview of Sec.~\ref{sec:disp_curves} on the interpretation and derivation of dispersion diagrams.\ This visualization serves as a user-guide while reading Sec.~\ref{sec:disp_curves}.}
	\label{fig:overview}
\end{figure*}

\section{Interpreting dispersion diagrams}
\label{sec:disp_curves}
This section provides a detailed discussion with graphical explanations on how dispersion diagrams are obtained using infinite periodic structure modeling, as commonly encountered in literature.\ 
A dispersion relation exists for the various wave types in the structure, cf.~Sec.~\ref{sec:background}, resulting often in a high amount of dispersion surfaces and corresponding curves, plotted in dispersion diagrams.\
To explain the derivation of the dispersion diagrams and to ease the interpretation of the resulting figures in this section, the same thin, flat isotropic plate of Sec. \ref{sec:background} is retaken, but only the bending wave propagation in this plate is considered while the longitudinal and shear waves are not considered.\
The bending waves can be isolated by modeling the plate with plate elements in the FE discretization, which only contain the DOFs related to out-of-plane motion.\ 

Fig.~\ref{fig:overview} gives an overview of the different aspects that are explained in this section.\ 
The visualization serves as a user-guide while reading the following section.\
The input is the dispersion eigenvalue problem of Eq.~(\ref{eq:EVP_BF}), and the section ends with the dispersion diagram.\
The section is divided into three parts: (i)~Sec.~\ref{subsec:disp_surf} introduces the dispersion surfaces and definition of the first Brillouin zone (FBZ), (ii)~Sec.~\ref{subsec:IBZ} discusses the transition from the FBZ to the irreducible Brillouin zone (IBZ), (iii)~Sec.~\ref{subsec:IBC} elaborates on the construction of the irreducible Brillouin contour (IBC) and the generation of the dispersion curves along it.\
In order to model the homogeneous plate with constant thickness of Sec.~\ref{sec:background} with infinite periodic structure theory, an arbitrary representative UC of dimensions $0.05 \times 0.05$~m is chosen.

\subsection{Dispersion surfaces and first Brillouin zone}
\label{subsec:disp_surf}
In Sec.~\ref{sec:background}, the dispersion surfaces were obtained analytically.\ 
However, for complex structures, the dispersion surfaces need to be computed numerically using for example the FE UC modeling.\ 
This section first uncovers how the numerically computed dispersion surfaces are obtained from the dispersion eigenvalue problem, how they look like and how they can be understood.\\

\begin{figure*}[h]
	\centering
	\includegraphics[width = 0.85\linewidth]{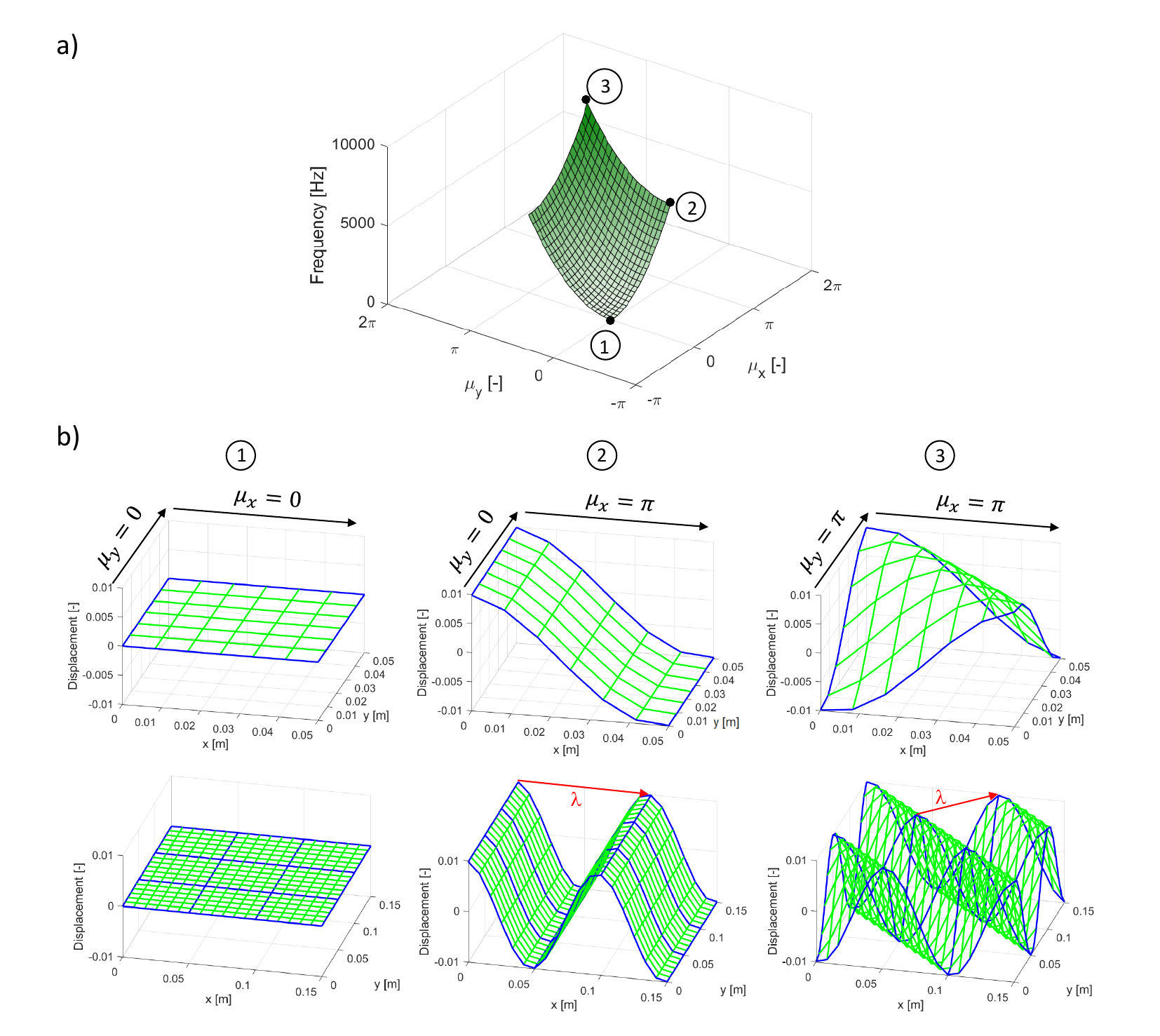}
	\caption{a) First bending wave dispersion surface with b) wave mode visualization for the points \textcircled{1}~$\rightarrow (\mu_x,\mu_y)=(0,0)$, \textcircled{2} $\rightarrow (\mu_x,\mu_y)=(\pi,0)$, and \textcircled{3} $\rightarrow (\mu_x,\mu_y)=(\pi,\pi)$ in the wave space.\ The wave mode is visualized for one UC as well as for a grid of $3\times 3$ UCs.}
	\label{fig:DC1}
\end{figure*}

\begin{figure*}
	\centering
	\includegraphics[width = \linewidth]{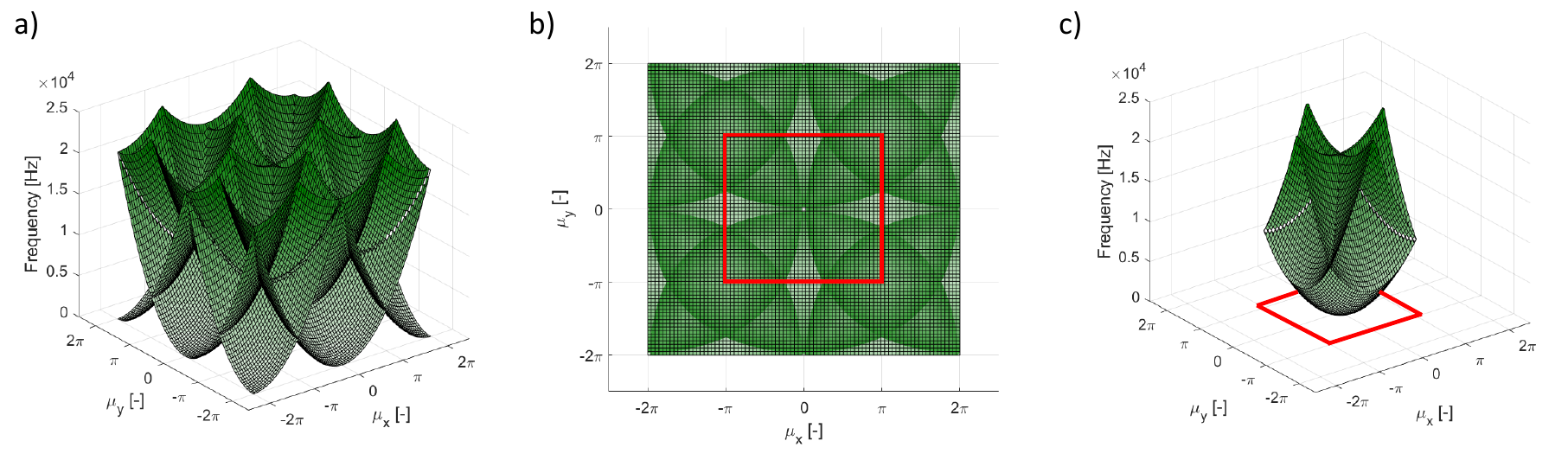}
	\caption{a) Dispersion surfaces for the first three solutions for imposed propagation constants in the domain $[-2\pi \hspace{0.1cm},\hspace{0.1cm} 2\pi] \times [-2\pi \hspace{0.1cm},\hspace{0.1cm} 2\pi]$, b) Top view of the dispersion surfaces of figure a, c) First three dispersion surfaces within the FBZ $[-\pi \hspace{0.1cm},\hspace{0.1cm} \pi] \times [-\pi \hspace{0.1cm},\hspace{0.1cm} \pi]$. 
 }
	\label{fig:DC2}
\end{figure*}

\begin{figure*}[t]
	\centering
	\includegraphics[width = \linewidth]{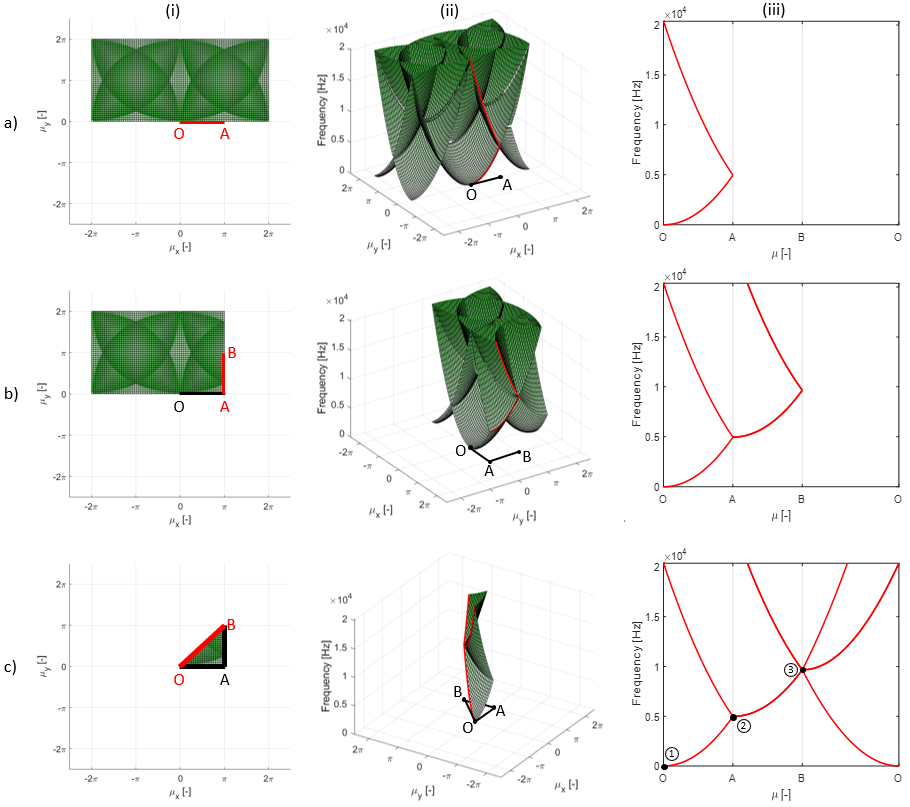}
	\caption{Graphical derivation to go from the dispersion surfaces in the IBZ to the dispersion diagram along the IBC. (i) Top view of the dispersion surfaces, (ii) 3D  view of the dispersion surfaces, (iii) dispersion curves, for respectively, a)~section OA, b)~section AB and c)~section OB of the dispersion diagram.}
	\label{fig:DC_final}
\end{figure*}

Using the inverse approach, real wave propagation constants ($\mu_x,\mu_y$) are imposed to the eigenvalue problem, cf.~Eq.~(\ref{eq:EVP_BF}) and solved for real frequencies resulting in dispersion surfaces.\
Fig.~\ref{fig:DC1}a shows part of the first bending wave dispersion surface 
and is obtained by plotting the smallest eigenvalue for each of the imposed ($\mu_x,\mu_y$)-pairs covering the region for $\mu_x$ ranging from $0$ till $\pi$ and $\mu_y$ ranging from $0$ till $\pi$, further symbolically denoted as $[0\hspace{0.1cm},\hspace{0.1cm} \pi] \times [0\hspace{0.1cm},\hspace{0.1cm}\pi]$.\ 
Note that this numerically obtained dispersion surface corresponds to the dispersion surface for freely propagating bending waves of Fig.~\ref{fig:an_DC}.\ 
In Fig.~\ref{fig:DC1}, however, the axes are expressed in terms of the wave constants ($\mu_x,\mu_y$) rather than the wave numbers ($k_x,k_y$), and the dispersion relation is only visualized for one quadrant in the wave constant domain ($\mu_x,\mu_y$), whereas in Fig.~\ref{fig:an_DC} the dispersion relation is visualized upto a fixed frequency.\
The reason for choosing this specific region in the wave constant domain will become clear during the discussion of Fig.~\ref{fig:DC2} and Fig.~\ref{fig:DC_final}.\\ 

The resulting surface, depicting frequencies in the wave space ($\mu_x,\mu_y$), indicates at which frequencies a bending wave can freely propagate through the structure and what the corresponding phase shift per UC will be, being $\mu_x$ respectively, $\mu_y$ when traveling in the $\mathbf{d}_x$, respectively, $\mathbf{d}_y$ direction.\ 
Further insights in the nature of the wave propagation can be gained by also considering the resulting eigenvectors, which correspond to the wave modes.\
Fig.~\ref{fig:DC1}b visualizes the wave modes corresponding to the points \textcircled{1}~$\rightarrow (\mu_x,\mu_y) = (0,0)$, \textcircled{2} $\rightarrow (\mu_x,\mu_y) = (\pi,0)$, and \textcircled{3} $\rightarrow (\mu_x,\mu_y) = (\pi,\pi)$.\
For each of these points the wave mode is visualized in both one UC as well as in a grid of $3\times 3$ UCs.\
The wavelength ($\varlambda$) can also be determined for each solution: e.g., for point \textcircled{2}, the wavelength equals $2L_x$, since $\mu_x = \pi = k_x L_x$ and $\varlambda = 2\pi/k_x$ in this case (as $\mu_y=0)$.\
Physically, $\mu_x = \pi$ means that half a wavelength is imposed over one UC, so a full wavelength covers two UCs.\ 
In Fig.~\ref{fig:DC1}b, the wave propagation direction and wavelength are shown.\\

Although only one part of the bending wave dispersion surface is shown in Fig.~\ref{fig:DC1}, solving Eq.~(\ref{eq:EVP_BF}) will of course lead to multiple solutions for each ($\mu_x,\mu_y$)-pair.\ 
For example, when solving for the first three solutions for each ($\mu_x,\mu_y$)-pair, three dispersion surfaces of increasing frequencies are obtained.\ 
Fig.~\ref{fig:DC2}a shows these three dispersion surfaces 
when solving ($\mu_x,\mu_y$) for the range $[-2\pi \hspace{0.1cm},\hspace{0.1cm} 2\pi] \times [-2\pi \hspace{0.1cm},\hspace{0.1cm} 2\pi]$, with a top view in Fig.~\ref{fig:DC2}b.\

It can be seen that the solutions are  $2\pi$-periodic in both $\mu_x$ and $\mu_y$  and that the dispersion surfaces have multiple intersections.\ 
In fact, due to the periodicity of the wave solution, as indicated by Bloch's theorem, also the dispersion surfaces are periodic in the wave space and have periodicity directions which are defined through the periodicity directions of the periodic structure~\cite{brillouin1946wave}.\
As a consequence, the dispersion solutions do not need to be calculated for the infinite set of possible ($\mu_x,\mu_y$)-pairs, but a confined zone can be defined that entails all dispersion information.\ 
Brillouin~\cite{brillouin1946wave} determined a method for constructing periodicity zones for various periodicity directions, and these zones were later named Brillouin zones.\ 
The first Brillouin zone (FBZ) for this case with orthogonal periodicity directions is the domain $[-\pi \hspace{0.1cm},\hspace{0.1cm} \pi] \times [-\pi \hspace{0.1cm},\hspace{0.1cm} \pi]$, surrounded by the red square in Fig.~\ref{fig:DC2}.\ 
Hence, all wave dispersion information is contained in Fig.~\ref{fig:DC2}c~\cite{brillouin1946wave}.\\

The periodicity of the dispersion surfaces has been proven by Brillouin~\cite{brillouin1946wave}.\
Here, only an intuitive explanation of the proof is given, supported by a graphical representation.\

When calculating the dispersion surfaces, the function $\omega(\boldsymbol\mu)$ is obtained through enforcing a relationship between the UC boundary DOFs in terms of  multiplications with $\lambda_x = e^{\mathrm{i}\mu_x}$ and/or $\lambda_y = e^{\mathrm{i}\mu_y}$.\ 
This relationship is, however, insensitive to a shift of the wave propagation constants $\mu_x$ or $\mu_y$ with a multiple of $2\pi$, e.g.~$\lambda_x = e^{\mathrm{i}\mu_x} = e^{\mathrm{i}(\mu_x +2\pi)}$.\ 
As a result, a shift with a multiple of $2\pi$  in the wave domain along either  $\mu_x$ or $\mu_y$, will lead to exactly the same solution of the dispersion relations, and the dispersion surfaces thus are $2\pi$-periodic in both $\mu_x$ and $\mu_y$.

\begin{figure*}[h]
	\centering
	\includegraphics[width = \linewidth]{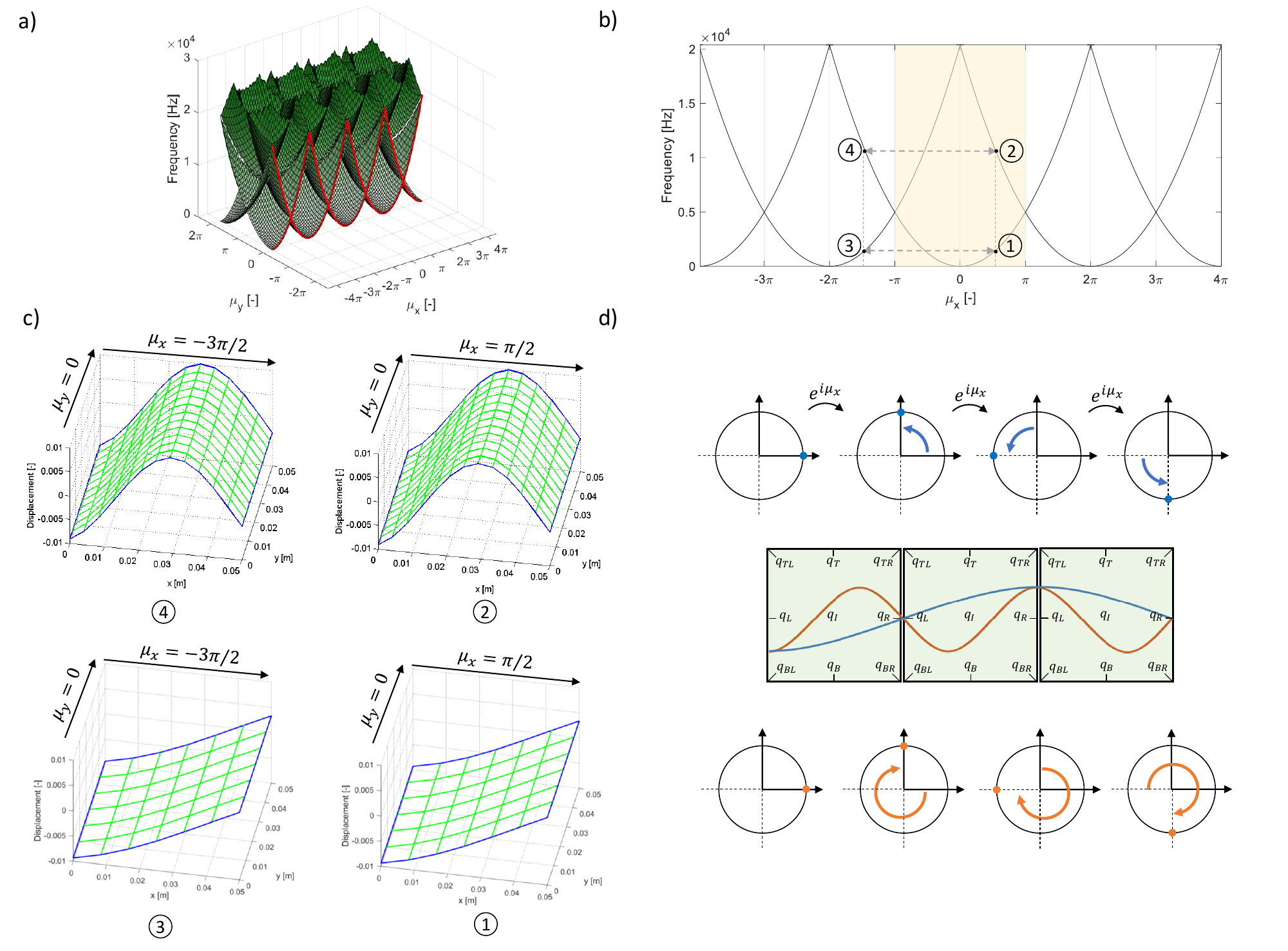}
	\caption{Visualization of periodicity in the wave domain.\ a)~Dispersion surfaces over the domain $[-4\pi \hspace{0.1cm},\hspace{0.1cm} 4\pi] \times [0 \hspace{0.1cm},\hspace{0.1cm} 2\pi]$, b)~2D cross section of the dispersion surfaces for $\mu_y=0$ with in yellow the FBZ, c)~Wave modes corresponding to the points in subfigure b, d)~1 dimensional representation of the out-of-plane wave mode amplitudes and phase shifts across each UC for point \textcircled{1} and \textcircled{3} (blue) and point \textcircled{2} and \textcircled{4} (orange).}
	\label{fig:BZ_fig}
\end{figure*}

To illustrate this in more detail, the pairs of propagation constants $(\mu_x,\mu_y)=(\pi/2,0)$ and $(\mu_x,\mu_y)=(\pi/2-2\pi,0)=(-3\pi/2,0)$ are considered in Fig.~\ref{fig:BZ_fig}.\ 
Both pairs lie on the $(\omega,\mu_y=0)$-plane and hence for this analysis the diagram of Fig.~\ref{fig:BZ_fig}b is used, which is the result of the evaluation of the eigenvalue problem for $\mu_y=0$.\ 
Visually this diagram can be interpreted as a cross section of the earlier obtained dispersion surface along the $(\omega,\mu_y=0)$-plane, as shown in Fig.~\ref{fig:BZ_fig}a, for which the red curves indicate the solutions for $\mu_y=0$.\
One can notice, again, the $2\pi$ periodicity in the diagram of Fig.~\ref{fig:BZ_fig}b: at each $2\pi$-shift with respect to $\mu_x=0$, a pair of parabolic dispersion curves originates and propagates in positive (right-going wave) and negative (left-going wave) direction.\

\begin{itemize}
    \item When solving the eigenvalue problem for $(\mu_x,\mu_y)=(\pi/2,0)$, one finds point \textcircled{1} and \textcircled{2} as the first two solutions.\
Point \textcircled{1} belongs to the dispersion curve starting in the positive direction with as origin $(\mu_x,\mu_y)=(0,0)$, while it can be seen that point \textcircled{2} belongs to the dispersion curve starting in the negative direction with as origin $(\mu_x,\mu_y)=(2\pi,0)$.\

\item When solving the eigenvalue problem for $(\mu_x,\mu_y)=(-3\pi/2,0)$, one finds points \textcircled{3} and~\textcircled{4} as the first two solutions.\
Point~\textcircled{3} is located on the dispersion curve starting in the positive direction with origin $(\mu_x,\mu_y)=(-2\pi,0)$, while point \textcircled{4} is located on the dispersion curve starting in the negative direction with as origin $(\mu_x,\mu_y)=(0,0)$.\

\end{itemize}

Figure~\ref{fig:BZ_fig}c shows the wave mode within one UC for the four derived solutions \textcircled{1}, \textcircled{2}, \textcircled{3} and \textcircled{4} together with the corresponding imposed propagation constants.\
Clearly,  solution \textcircled{1} and \textcircled{3}, resp. \textcircled{2} and \textcircled{4} are identical, and solving only for one of both propagation constants would suffice.
When inspecting the wave numbers of the wave modes, intuitively, one might have expected to only find solution \textcircled{1} for propagation constants $(\mu_x,\mu_y)=(\pi/2,0)$ and solution \textcircled{4} for propagation constants $(\mu_x,\mu_y)=(-3\pi/2,0)$ since this would correspond with the wave numbers of the corresponding wave modes.\ 
However, since only relations between the boundaries are enforced by Bloch's theorem, i.e.~$\lambda_j = e^{\mathrm{i}\mu_j} = e^{\mathrm{i}(\mu_j +2\pi)}$ with $j=x,y$, one finds identical solutions for $(\mu_x,\mu_y)=(\pi/2,0)$ and $(\mu_x,\mu_y)=(-3\pi/2,0)$.\ 
Figure~\ref{fig:BZ_fig}d schematically explains this further.\ 
In this figure, three consecutive UCs in the $x$-direction are shown and the wave mode for \textcircled{1} (or  \textcircled{3}, as they are identical) is represented in 2D with a full blue line whereas the wave mode for \textcircled{2} (or \textcircled{4}) is represented with a full orange line.\ 
Above and below the UCs, the phase shifts that occur at each UC boundary are indicated for, respectively, the blue and orange curve.\ 
Although the wave amplitude within each UC is different, it is clear that, when only inspecting the phase shift at the boundaries, no distinction can be made between a phase shift $\pi/2$  per UC, i.e., $\mu_x=\pi/2$, and a phase shift $-3\pi/2$ per UC, i.e,  $\mu_x=-3\pi/2$, or any other phase shift $\mu_x= \pi/2 \pm 2\pi n$ with $n \in \mathbb{N}$, for that matter.\  
This shows that dispersion surfaces exhibit a $2\pi$-periodicity, that is $\omega(\mu_x,\mu_y) =  \omega(\mu_x \pm 2\pi m, \mu_y \pm 2\pi n)$ with $m,n \in \mathbb{N}$, and hence, the information in the FBZ ($[-\pi \hspace{0.1cm},\hspace{0.1cm} \pi] \times [-\pi \hspace{0.1cm},\hspace{0.1cm} \pi]$) indeed contains all the necessary unique dispersion surface results.

\begin{figure*}
	\centering
	\includegraphics[width = 0.8\linewidth]{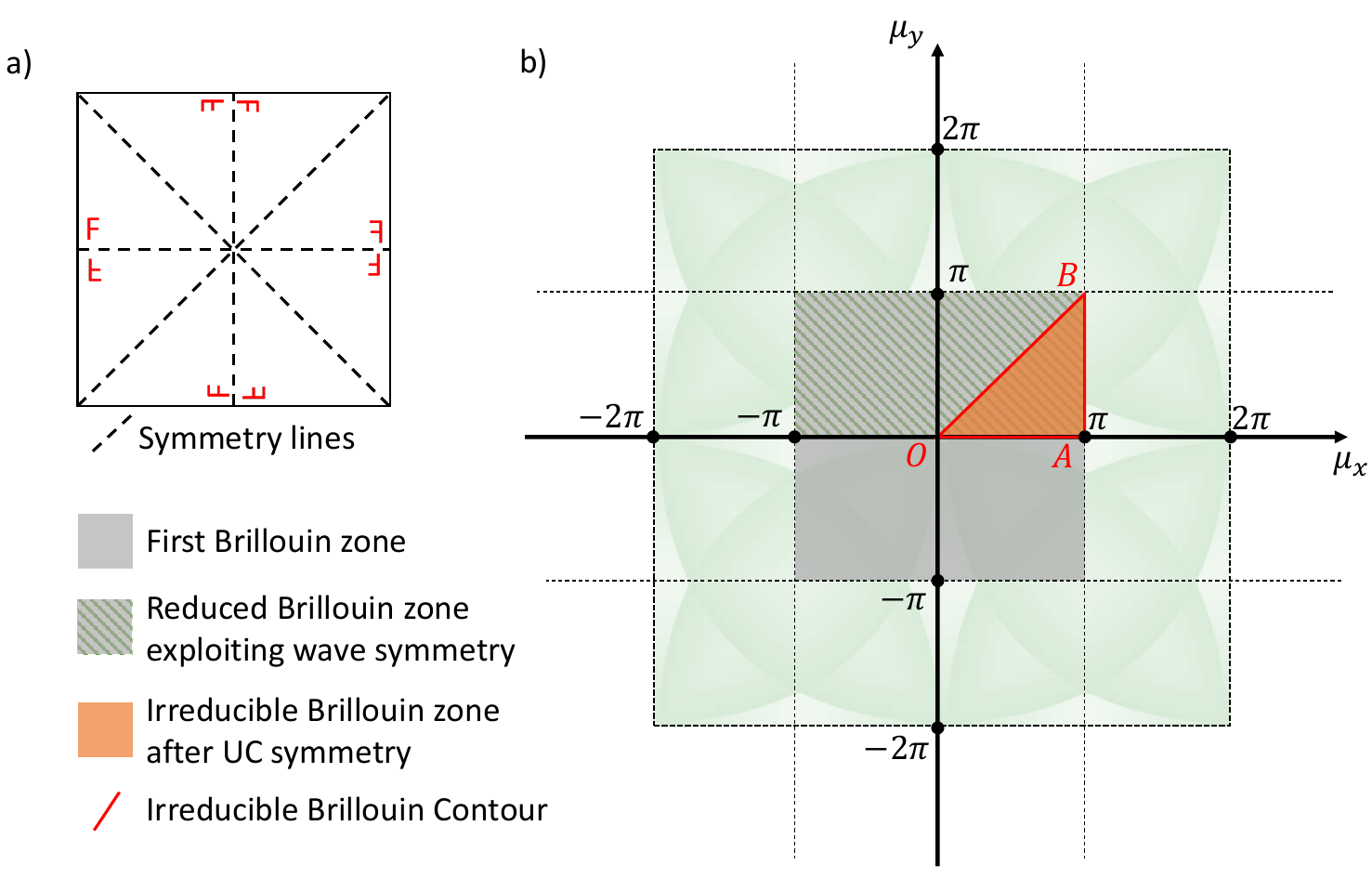}
	\caption{Example UC belonging to plane crystallographic group p4mm (with 4 symmetry lines) together with a top view of the dispersion surfaces with visualization how to obtain the IBC.}
	\label{fig:DC3}
\end{figure*}

\subsection{Irreducible Brillouin zone}
\label{subsec:IBZ}
Although the FBZ ($[-\pi \hspace{0.1cm},\hspace{0.1cm} \pi] \times [-\pi \hspace{0.1cm},\hspace{0.1cm} \pi]$) was shown to contain all dispersion information, the wave domain in which the dispersion eigenvalue problem needs to be solved can even be further reduced by exploiting symmetry in (i) wave propagation and (ii) UC geometry.\ 
These symmetries lead to symmetries in the wave domain and result in a so-called Irreducible Brillouin zone (IBZ).\ 
Due to symmetry, dispersion surfaces only need to be calculated in the IBZ to characterize the wave propagation in the infinite periodic structure.\ 
This is illustrated in Fig.~\ref{fig:DC3}b for the here considered highly symmetric UC example of Fig.~\ref{fig:DC3}a.\ This particular UC contains 4 lines of symmetry and belongs to the plane crystallographic group (PCG) p4mm\footnote{A plane crystallographic group, also denoted as wallpaper group, is a topological concept which describes several symmetry patterns in two-dimensional lattices. The naming p4mm denotes that the considered UC contains a 4-fold rotational symmetry, i.e.~90° rotational symmetry, while two mirror symmetry axes in the perpendicular direction are present.}~\cite{maurin2018probability}.\ 
In general, the shape of the IBZ depends on the type of symmetries present in the UC.\ 
To grow a better understanding of how the IBZ is obtained, its construction is next discussed in more detail for the here considered case in two steps: including (i) the wave propagation symmetry and (ii) the symmetry of the UC geometry.\\

\begin{figure*}
	\centering
	\includegraphics[width = 0.7\linewidth]{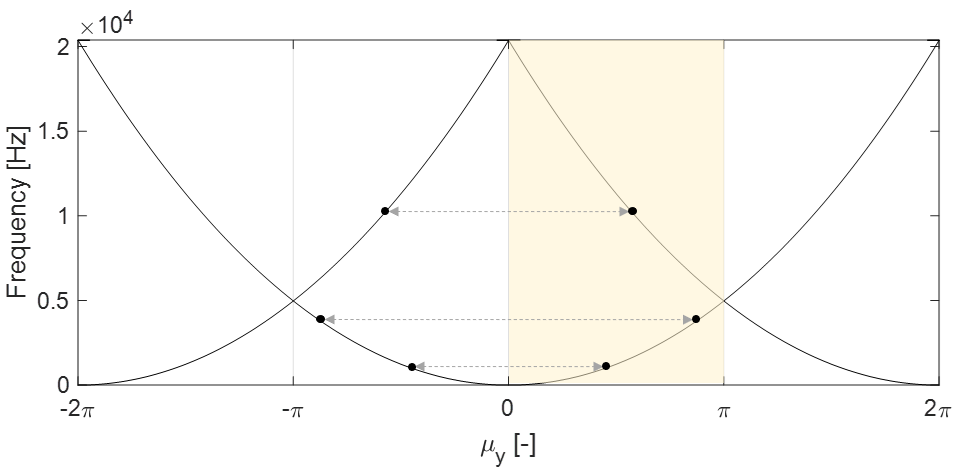}
	\caption{Visualization to show the symmetry between the left and right going waves. A dispersion diagram evaluated for  $\mu_x=0$ with connection between right going ($\mu_y = \beta$) and left going ($\mu_y = -\beta$) waves. The resulting reduced Brillouin zone is indicated in yellow.}
	\label{fig:DC_symm}
\end{figure*}

(i)~For time-independent linear systems, the properties of a wave propagating along an axis do not depend on the sense of propagation~\cite{brillouin1946wave}.\ 
Hence, the dispersion surfaces are even functions and only half the FBZ needs to be solved to obtain all dispersion information, visualized in Fig.~\ref{fig:DC_symm}.\ 
Although this reduction can be made along any axis of choice, in this paper it is chosen to use the $y$-axis and hence reduce the FBZ to the zone ($[-\pi \hspace{0.1cm},\hspace{0.1cm} \pi] \times [0 \hspace{0.1cm},\hspace{0.1cm} \pi]$).\\

(ii)~In case the UC possesses no internal geometrical symmetries, the derived reduced Brillouin zone in the previous step will be the IBZ.\ 
In case of internal symmetries, the reduced Brillouin zone can be further reduced to its IBZ.\ 
Based on the type and combination of symmetries (rotational, glide and mirror symmetry), a classification can be made on the resulting shape of the IBZ~\cite{maurin2018probability,cracknell1974tables}.\
In this work, the example of the highly symmetric UC of Fig.~\ref{fig:DC3}a, belonging to PCG p4mm, is considered and the corresponding IBZ is highlighted in orange in Fig.~\ref{fig:DC3}b  and is defined as the zone surrounded by the wave propagation constants O~$\rightarrow (0,0)$, A~$\rightarrow (\pi,0)$ and B~$\rightarrow (\pi,\pi)$.\

\begin{figure*}
	\centering
	\includegraphics[width = \linewidth]{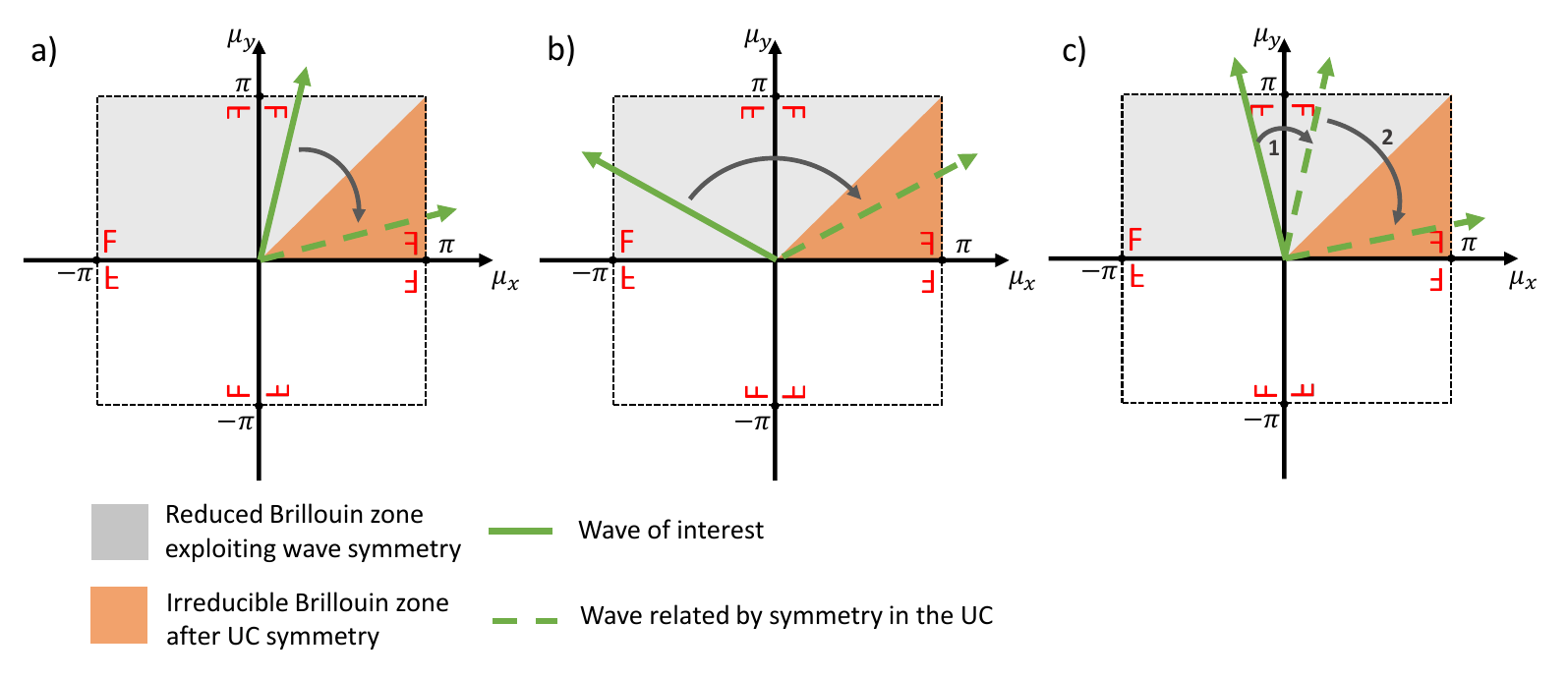}
	\caption{Visualization to illustrate that the IBZ (orange) contains all the required information from within the FBZ.\ For each wave (solid green) outside the IBZ, a corresponding wave within the IBZ (dashed green) can be found using the geometrical symmetry of the UC (visualized with the red 'F' symbols).}
	\label{fig:DC5}
\end{figure*}

To understand how this IBZ allows representing all information in the FBZ, three cases of wave propagation are considered in Fig.~\ref{fig:DC5}.\ 
Each time it will be shown how the considered wave propagation direction, represented by a solid green arrow, can be related to a wave propagation direction studied within the IBZ, represented by a dashed green arrow.\ 
Note that in this figure only the direction of the wave is defined, as the argumentation only depends on the ratio and sign of the wave propagation constants $(\mu_x,\mu_y)$ and not on their absolute amplitude.\ 
To better highlight the symmetry of the UC, multiple 'F'-symbols at various places and orientations are added to the schematic representation of the UC (cf.~Fig.~\ref{fig:DC3}). This resulting representation reflects the PCG p4mm, which can be mirrored around $x$, $y$, and its diagonals, and which is the symmetry of the UC studied in this section. 

\begin{itemize}
    \item Fig.~\ref{fig:DC5}a shows the case for a wave in the first quadrant but not in the IBZ.\ Due to the diagonal symmetry of the UC (Fig.~\ref{fig:DC3}a), this wave will contain the same information as the wave represented by the green dotted arrow in the IBZ.\
    \item Fig.~\ref{fig:DC5}b shows the case for a wave in the second quadrant.\ Due to the vertical symmetry of the UC, this wave will contain the same information as the wave represented by the green dotted arrow in the IBZ.\ 
    \item  Fig.~\ref{fig:DC5}c shows again a case for a wave in the second quadrant.\ In this case, the green arrow lies in the first quadrant after applying vertical symmetry (step 1), but not yet in the IBZ.\ Therefore, the diagonal symmetry needs to be applied as well (step 2), after which the projection will end up in the IBZ.\ After these two steps, it can be seen that the original arrow will contain the same information as the wave represented by the green dotted arrow in the IBZ.\
\item Lastly, cases could be considered for which the waves lie in the third or fourth quadrant.\ Each of these cases could be mapped on one of the previously treated cases by considering the symmetry in the wave propagation.\ For this UC, which also is symmetric around the $y$-axis, the horizontal symmetry could also be applied instead of the symmetry in the wave propagation.  
\end{itemize}
This illustrates that the IBZ contains all unique information of the FBZ.\
For a more detailed discussion on UCs with other symmetries, the reader is referred to~\cite{maurin2018probability}.\

\subsection{Irreducible Brillouin contour}
\label{subsec:IBC}
At this point, solving the dispersion eigenvalue problem for imposed wave propagation constants in the IBZ still results in a series of dispersion surfaces, which can take considerable computation time and is not always clear to interpret.\ 
Often, one is not interested in the complete dispersion surfaces, but, e.g. only in the minima and maxima on each dispersion surface, as will be shown in an example later.\

For this reason, the dispersion surfaces are often only evaluated along the boundary of the IBZ (Fig.~\ref{fig:DC3}), named the irreducible Brillouin contour (IBC).\ 
This approach is based on the common assumption that all minima and maxima of the dispersion surfaces over the IBZ occur on the boundaries of this IBZ.\ 
Although no explicit proof exists, Maurin et al.\ gave a stochastic proof~\cite{maurin2018probability}.\ 
Intuitively the evaluation of the solutions on the IBC to find extremes on a dispersion surface can be motivated by the fact that this contour connects the wave propagation constants with a phase difference of $\pi$ which are the solutions for which (multiple) half wavelengths occur across a UC.\
These points correspond to solutions with a high chance of possible wave interference which enable interesting phenomena, as discussed later.\ 

The remainder of this section gives a visual interpretation of the relationship between the dispersion surfaces in the wave space and the dispersion diagram along the IBC.\
This with the goal to ease the interpretation of the dispersion diagrams which result from evaluating the eigenvalue problem of Eq.~(\ref{eq:EVP_BF}) along the IBC.\
For the highly symmetric UC belonging to PCG p4mm (Fig.~\ref{fig:DC3}a), the IBC is denoted by the OABO trajectory consisting of wave propagation constants~($\mu_x,\mu_y)$:
\begin{equation}
    (0,0) \longrightarrow (\pi,0) \longrightarrow (\pi,\pi) \longrightarrow (0,0).
\end{equation}

The dispersion diagram along this contour can be envisioned as subsequent cuts that are made between the dispersion surfaces and planes along the OA, AB and BO edges of the IBC.\
The resulting intersecting curves for OA, AB and BO are next piece-wise plotted on a 2D plot of frequency versus wave propagation along this OABO trajectory, effectively establishing the dispersion diagram along the IBC.\
This process is visualized in Fig.~\ref{fig:DC_final} for each of the sections OA, AB and BO along the IBC, with the resulting dispersion curves shown in the dispersion diagram in Fig.~\ref{fig:DC_final}c.iii.\
The points \textcircled{1}, \textcircled{2} and \textcircled{3} from Fig.~\ref{fig:DC1}a are indicated in the final dispersion diagram as well.\ 
Hence, it can be seen that each point on the dispersion curve represents a certain wave propagating freely through the infinite periodic structure along the direction corresponding to the imposed wave propagation constants, and at a certain frequency.\ 

In conclusion, the preceding paragraphs provided an in-detail description of how the evaluation of infinitely large dispersion surfaces can be limited to the evaluation of dispersion diagrams which capture the main information needed to understand the wave propagation in an infinite periodic structure.\ 
Of course, in practice these dispersion diagrams are computed directly by solving the eigenvalue problem~Eq.~(\ref{eq:EVP_BF}) only for wave propagation constants ($\mu_x,\mu_y$) along the IBC.\
In the next section, a \textsc{Matlab} implementation is discussed which performs exactly this calculation.\

\section{Matlab implementation for dispersion curve calculation}
\label{sec:Matlab_impl_new}
Having established a basic background in calculating dispersion curves and interpreting dispersion diagrams in the preceding sections, this section elaborates on a \textsc{Matlab} implementation to compute these dispersion curves.\ 
The \textsc{Matlab} code is provided in \ref{app:matlab} (and the function \texttt{FEM\_incompatible\_modes} in the \textit{Supplementary Materials}).\ 
The corresponding \textsc{Matlab} scripts are available on the GitHub repository {\url{https://www.github.com/LMSD-KULeuven/2D\_InverseUndamped\_DispersionCurves}}.\
\newline
This section serves as a guide on how to use the implementation and discusses three initial case studies.
The code is written from a didactical point of view, with the premise of readability and understandability, not computational efficiency.\
The interested reader is referred to \ref{sec:matlab_impl} for a detailed discussion on the \textsc{Matlab} implementation and to \ref{sec:ext} for a discussion on possible extensions of the code.\

As discussed in the introduction of Sec.~\ref{sec:disp_curves}, plate elements are used to model the plate in that section in order to obtain only bending waves and ease the interpretation of the resulting dispersion surfaces.\
For the \textsc{Matlab} implementation discussed in this section, the choice is made to use linear elastic hexahedral solid elements to discretize the plate under investigation.\
This results in a more generally applicable code, representing all wave types (bending, longitudinal, and shear), and allows extensions to all possible topologies and eases adaptation towards 3D periodicity.\\

\begin{figure*}
	\centering
	\includegraphics[width = \linewidth]{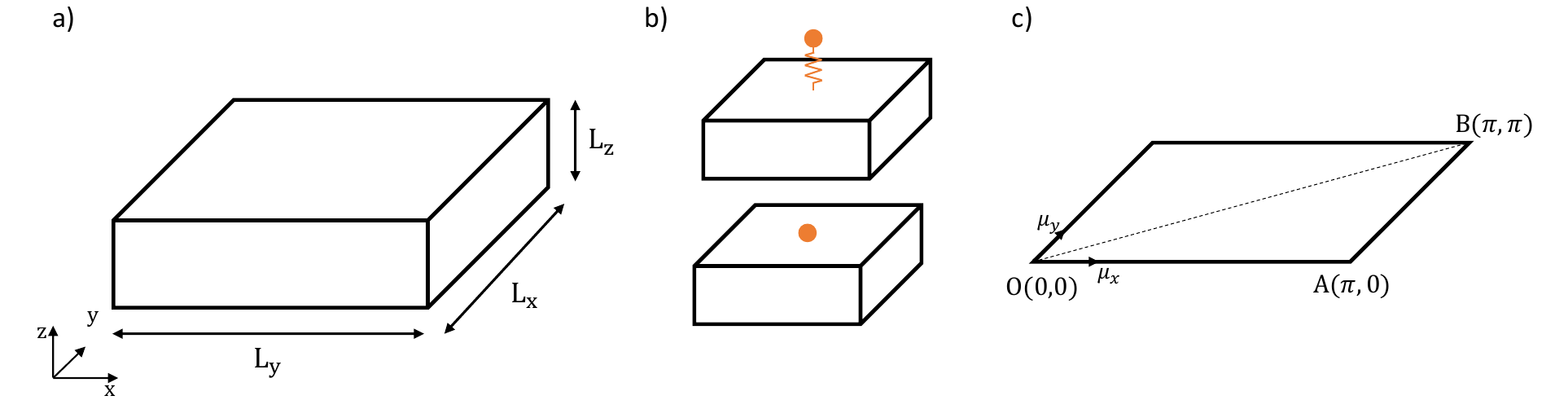}
	\caption{Schematic overview of the inputs. a)~UC definition. b)~Options for resonator or mass addition to the UC. c)~IBC convention.}
	\label{fig:inputs}
\end{figure*}

\subsection{Input data (lines 4-25)}
The first section of the code collects all the required user inputs which define the problem.\
An overview for a specific example is given in Tab.~\ref{tab:inputs}.\
A 2D periodic UC is considered which consists of a plate with dimensions \verb+Lx+, \verb+Ly+, \verb+Lz+ in the $x$-, $y$- and $z$-direction, respectively (Fig.~\ref{fig:inputs}a).\
The plate is assumed to be composed of a linear isotropic elastic material with Young's modulus \verb+E+~[Pa], density \verb+rho+~[kg/m$^3$] and Poisson's ratio \verb+v+~[-].\
The plate is discretized with solid elements, while the mesh is defined by the variables \verb+n_elx+, \verb+n_ely+, \verb+n_elz+, which represent the number of elements in the $x$-, $y$- and $z$-direction, respectively.\
Note that in the code, the FE-part is hard-coded for hexahedral solid elements with $3$ DOFs per node, and which are all of equal size.\
Generally speaking, however, a UC FE mesh can of course be composed of elements of different shapes and sizes.\ 
The main requirement for the FE mesh is that the nodes on the opposing sides of the UC (left-right, top-bottom and corners) correspond such that the periodic boundary conditions can be applied, cf.~Eqs.~(\ref{eq:Proj}).\ 
Using equal sized elements simplifies the FE implementation and satisfies this requirement.\

The possibility to introduce a lumped scatterer in the center of the UC is foreseen in order to obtain a truly periodic structure with associated interesting effects on the dispersion behavior.\ 
To this end, either a mass-spring resonator with a $z$-direction displacement DOF or a point mass can be added to the middle of the UC's top face (Fig.~\ref{fig:inputs}b) by changing the variable \verb+scatterer+, to \verb+resonator+ or \verb+mass+, respectively.\
If one is interested in computing the dispersion curves of the plate only, \verb+scatterer+ should equal \verb+none+.\
With \verb+f_res+~[Hz] and \verb+m_ratio+~[-] the added scatterer can be tuned: \verb+m_ratio+ defines the relative mass of the resonator or point mass, as compared to the plate mass, \verb+f_res+ defines the frequency to which the mass-spring system is tuned and is relevant when \verb+scatterer+ equals \verb+resonator+.

Before proceeding, the IBC should be defined.\
Fig.~\ref{fig:inputs}c gives the convention which is followed for the 2D periodic case belonging to PCG p4mm as visualized in Fig.~\ref{fig:DC3}a.\
Note that when an uneven number of elements \verb+n_elx+, \verb+n_ely+ is chosen, a possible scatterer will not be placed  exactly in the middle of the plate's top face so the symmetry will be lost and the IBC needs to be adapted correspondingly.\
The variable \verb+cont_name+ defines the coordinates names of the IBC, which are later used for plotting.\
The IBC coordinates,  consisting of a matrix with the wave propagation constants ($\mu_x$, $\mu_y$) per row, are defined by the variable \verb+cont_co+ and are used for the calculation of the dispersion diagrams.
For example, when defining the IBC as OABO, the code reads as follows:
\begin{lstlisting}[columns=flexible , numbers=left, firstnumber=22, style=Matlab-editor, basicstyle=\small]
cont_name = {'O','A','B','O'};  
cont_co =  [0,0; pi,0; pi,pi; 0,0]; 
\end{lstlisting}
The user can also choose the resolution with which the dispersion curves will be computed, given by the variable \verb+step_size+.\ 
A smaller \verb+step_size+ results in a finer resolution, but also a higher computation cost.\
Finally, the variable \verb+n_curves+ represents the number of dispersion curves that are computed per calculation point on the IBC.\

\subsection{Discussion results}
\label{subsec:results}
Using the \textsc{Matlab} code, this section aims to discuss some results for the specific case of a plate, with a possible addition of a point mass or a mass-spring resonator.\ 
These cases are selected based on the fact that such additions lead to interesting changes to the dispersion behavior of the structure, relevant for the educational purpose of this paper.\ 
While it is not our intention to provide the reader with an in-depth discussion of the physics associated to such structures, we briefly explain the effect of these additions.\ 
As mentioned in the introduction, periodic structures can exhibit stop bands, which are frequency zones in which freely propagating waves are inhibited.\ 
By adding periodic scatterers, e.g.\ point mass additions, to a host structure, e.g.\ a plate, a phononic crystal can be obtained, see e.g.~\cite{sigalas1993band}.\ 
The stop bands in those structures rely on so-called Bragg scattering, where destructive interference occurs between reflected and transmitted waves, and whereby the stop band frequencies are directly linked to the length scale of periodicity.\ 
By adding mechanical resonators to a host structure on a subwavelength scale, stop bands are obtained due to Fano-type interference between the incoming waves and the waves re-radiated by the resonant cells, see e.g.~\cite{goffaux2002evidence}.\ 
The corresponding stop band frequency range is driven by the subwavelength resonators' tuned frequency.\ 
Hence, these stop bands do not depend on the periodicity length scale and can be achieved at comparatively lower frequencies without requiring a large spacing, using scales much smaller than the wavelength.

\begin{table}
    \centering
    \begin{tabular}{ |c|c|c|c| } 
    \hline
    \textbf{UC size} [m] & \textbf{Material} & \textbf{Discretization} & \textbf{IBC definition} \\
    \hline
    \verb+Lx+ $ = 0.05$ & \verb+E+ $= 210\mathrm{e}9$~Pa& \verb+n_elx+ $=10$ & \verb+cont_name+ = {O,A,B,O} \\
    \hline
    \verb+Ly+ $ = 0.05$ & \verb+v+ $ = 0.3$ & \verb+n_ely+ $=10$ & \verb+cont_co+ =  [0,0; pi,0; pi,pi; 0,0] \\
    \hline
    \verb+Lz+ $ = 0.005$ & \verb+rho+ $= 7800$~kg/m$^3$ & \verb+n_elz+ $=3$ & \verb+step_size+ = $0.01\pi$ \\
    \hline
     &  & \verb+n_dof+ $=3$ & \verb+n_curves+ $=10$ \\
    \hline
    \end{tabular}
    \caption{Specific inputs for the plate used for the results in Sec.~\ref{subsec:results}.}
    \label{tab:inputs}
\end{table}

\begin{figure*}
	\centering
	\includegraphics[width = \linewidth]{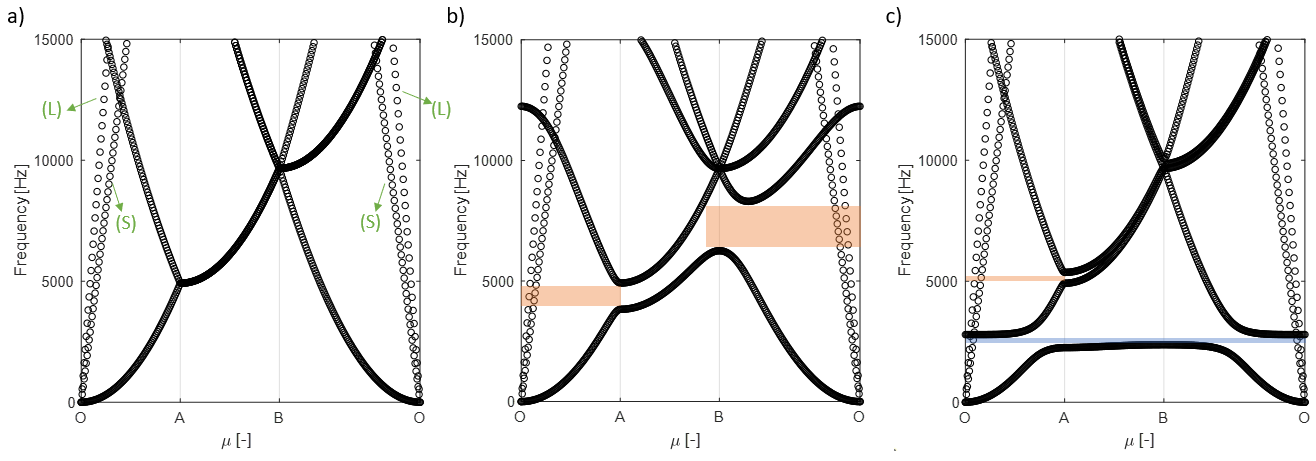}
	\caption{Results obtained with the educational \textsc{Matlab} code with inputs given in Tab.~\ref{tab:inputs}.\ Only the black parts are obtained with the code, while the colored parts are added in post-processing to help the interpretation.\ a)~Dispersion diagram of the plate with indication of the longitudinal (L) and shear (S) wave modes.\ b)~Dispersion diagram when a point mass with mass ratio $0.3$ is added.\ c)~Dispersion diagram when a mass-spring resonator is added with mass ratio $0.3$ and tuned frequency $2500$~Hz.\ In orange, the partial bandgaps for the bending waves are shown.\ In blue, the full omni-directional bandgap for the bending waves is shown.}
	\label{fig:results}
\end{figure*}

Tab.~\ref{tab:inputs} gives the selected input parameters describing the UC of the plate and the definition of the IBC.\
The point mass is defined by \verb+m_ratio+ $=0.3$ , while the mass-spring system is tuned to \verb+f_res+ $=2500$~Hz and the same \verb+m_ratio+ $=0.3$.\
Fig.~\ref{fig:results} shows the obtained dispersion diagrams.\
These results are briefly discussed in this paper; a detailed discussion for these cases can be found in \cite{claeys2013potential}.

Fig.~\ref{fig:results}a shows the dispersion diagram of the bare plate.\ 
This is the same case as discussed in Sec.~\ref{sec:disp_curves}.\
The diagram is obtained by inserting the inputs of Tab.~\ref{tab:inputs} on lines 4-25 of the \textsc{Matlab} implementation, running the code and limiting the $y$-axis in the plot to $15$~kHz.\
Note that all wave types, bending, longitudinal (L) and shear (S) waves are now obtained in the dispersion diagram, similar to the analytical predicted ones in Sec.~\ref{sec:background}.\

The frequency for which the bending wavelength in the infinite plate without periodic additions equals twice the UC length in $x$-direction can be determined analytically using Kirchhoff plate theory:
\begin{equation}
    f_{\varlambda/2} = \frac{2\pi}{(2L_x)^2}\sqrt{\frac{E L_z^2}{12(1-\nu^2) \rho }}.
\end{equation}
This corresponds with point A where the bending wave dispersion curve folds and is also the frequency for which Bragg interference occurs.\
For the plate discussed above this frequency equals $4933$~Hz, which corresponds to the numerically obtained result (Fig.~\ref{fig:results}a).\

The point mass and resonator are scatterers that are explicitly added to influence wave propagation.\ 
It can be noticed that the added scatterers will have a negligible impact on the longitudinal and shear waves in the shown frequency range in what follows.\
Fig.~\ref{fig:results}b shows the dispersion diagram of the plate with the addition of the point mass.\
The curves are obtained by changing the variable \verb+scatterer+ to \verb+mass+ while \verb+m_ratio+ $=0.3$ and rerunning the code.\
A first opening of the dispersion curves occurs exactly at the Bragg interference limit since there destructive interference is possible.\
The $f_{\varlambda/2}$ is therefore the minimum frequency at which phononic crystals will have (partial) bandgaps, as shown in orange.\
On higher frequencies, other (partial) bandgaps occur as well.\
For increasing mass ratios, the dispersion curves will open up going from partial bandgaps to full omni-directional bandgaps.\
The reader can easily check this by increasing the \verb+m_ratio+.\

Fig.~\ref{fig:results}c shows the result when instead a resonator, tuned to $2500$~Hz, is added.\
The curves are obtained by changing the variable \verb+scatterer+ to \verb+resonator+ while \verb+m_ratio+ $=0.3$ and \verb+f_res+ $=2500$~Hz and rerunning the code.\
Around the tuned frequency, a Fano-type bandgap opens up for the bending waves.\
Again, the frequency $f_{\varlambda/2}$ is important, as resonators have to be added on a subwavelength scale, and hence below this frequency, in order to enable a full omni-directional bandgap of this type \cite{claeys2013potential}.\ 
The reader can explore this by changing the \verb+f_res+ accordingly.\ 
Changing the  \verb+m_ratio+, on the other hand, will impact the width of the stop band.\ 
It can be noted that, since the resonant additions also act as periodic scatterers, a narrow directional bandgap can also be identified around~$f_{\varlambda/2}$.\\

\section{Conclusion}
\label{sec:concl}
Understanding dispersion diagrams and computing dispersion curves can be a daunting challenge for novice researchers.\ 
To address this, this manuscript provides a guide and accompanying basic \textsc{Matlab} code for dispersion curve computations.

First, an introduction into dispersion relations and wave propagation is given in order to introduce the basic concepts to explain dispersion diagrams.\
Thereafter, the basic theory behind infinite periodic structure modeling is given.\
The focus is put on the inverse approach for dispersion relation calculations of 2D periodic media.\
Next, insights are given with a graphically supported explanation on how to read, interpret and derive the dispersion diagrams.\ 
Finally, a guide is provided to start using the code together with three example cases.\ 
All steps in the code are explained in detail in appendix together with numerous extensions as inspiration to extend/adapt the code.\

The \textsc{Matlab} code to numerically calculate dispersion curves based on a FE UC description is available in \ref{app:matlab} of this paper, and can be downloaded on the Github repository ({\url{https://www.github.com/LMSD-KULeuven/2D\_InverseUndamped\_DispersionCurves}}).\ 
Users which suggest modifications, extensions or improvements to the code are invited to share them on the public Github repository or by contacting the authors.\

\section*{Acknowledgments}
The research of V.\ Cool (fellowship no.\ 11G4421N) is funded by a grant from the Research Foundation - Flanders (FWO).\ 
The Research Fund KU Leuven is gratefully acknowledged for its support.\

\appendix

\section{Bloch's theorem}
\label{app:Bloch}
This section gives more details on how Eq.~(\ref{eq:qk}) is obtained.\
A complete discussion on Bloch's theorem can be found in literature, e.g.~\cite{bloch1929quantenmechanik,brillouin1946wave}.\
In essence, Bloch's theorem governs the wave propagation in periodic media.\
When a time-harmonic wave propagates through the structure, the theorem states it can be written as:
\begin{equation}
\label{eq:qA_1}
    \mathbf{q}(\mathbf{r},k,\omega) =  \tilde{\mathbf{q}}(\mathbf{r},k) e^{\mathrm{i}\mathbf{k} \cdot \mathbf{r}} e^{\mathrm{i} \omega t},
\end{equation}
in which $e^{\mathrm{i}\mathbf{k} \cdot \mathbf{r}}$ is a plane wave and $\tilde{\mathbf{q}}$ is a periodic function which has a spatial periodicity due to the underlying periodicity of the structure:
\begin{equation}
\label{eq:qA_2}
     \tilde{\mathbf{q}}(\mathbf{r},k) = \tilde{\mathbf{q}}(\mathbf{r} + n_x \mathbf{d}_x + n_y \mathbf{d}_y,k),
\end{equation}
with $n_x$ and $n_y$ integer numbers denoting the UC positioning.\
By combining the above two equations, Eq.~(\ref{eq:qk}) is obtained, namely:
\begin{equation}
\label{eq:qA_3}
\begin{aligned}
    \tilde{\mathbf{q}}(\mathbf{r}_P,k,\omega) &= \tilde{\mathbf{q}}(\mathbf{r}_P,k) e^{\mathrm{i}\mathbf{k} \cdot \mathbf{r}_P} e^{\mathrm{i} \omega t} \\
    & = \tilde{\mathbf{q}}(\mathbf{r}_U,k) e^{\mathrm{i}\mathbf{k} \cdot \mathbf{r}_U} e^{\mathrm{i}\mathbf{k} \cdot ( n_x \mathbf{d}_x + n_y \mathbf{d}_y)} e^{\mathrm{i} \omega t} \\
    & = \mathbf{q}_{ref}(\mathbf{r}_U,k,\omega) e^{\mathrm{i}\mathbf{k} \cdot ( n_x \mathbf{d}_x + n_y \mathbf{d}_y)},
\end{aligned}
\end{equation}
in which first Eq.~(\ref{eq:qA_1}) is used while filling in $\mathbf{r}=\mathbf{r}_P$, next Eq.~(\ref{eq:qA_2}) and Eq.~(\ref{eq:rP}) are applied, afterwards again Eq.~(\ref{eq:qA_1}) is used.\

\begin{figure*}
	\centering
	\includegraphics[width = 0.5\linewidth]{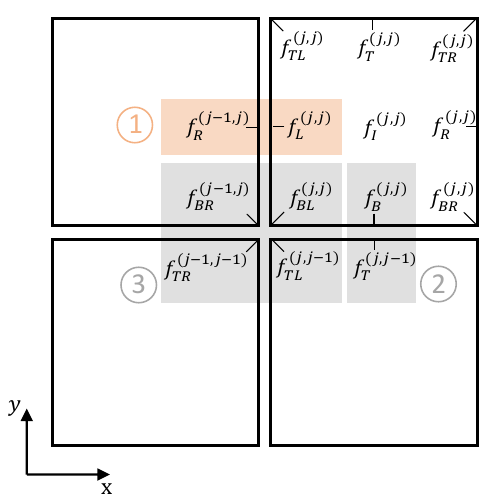}
	\caption{Equilibrium of the generalized forces at \textcircled{1} the left boundary, \textcircled{2} the bottom boundary and \textcircled{3} the corner written explicitly in which the subscript $(\mathrm{j,j})$ indicates it is the $\mathrm{j}^{th}$ cell in the $x$- and $y$-direction.}
	\label{fig:force_balance2}
\end{figure*}

\section{Force equilibrium}
\label{app:force}
This appendix further explains how Eqs.~(\ref{eq:force_bal}) are obtained.\
The reasoning is given here first for the left boundary to derive equation \textcircled{1} of Eqs.~(\ref{eq:force_bal}).\ 
Consider the force balance on the left boundary of the $(\mathrm{j,j})^{\mathrm{th}}$ cell, highlighted in orange on Fig.~\ref{fig:force_balance2}:
\begin{equation}\label{eq:force_app}
    \mathbf{f}_L^{(\mathrm{j,j})} + \mathbf{f}_R^{(\mathrm{j-1,j})} = 0.
\end{equation}
Using Bloch's theorem, the force $\mathbf{f}_R^{(\mathrm{j-1,j})}$ is related to the force $\mathbf{f}_R^{(\mathrm{j,j})}$ in the following manner: 
\begin{equation}
    \mathbf{f}_R^{(\mathrm{j,j})} = e^{\mathrm{i}\mu_x} \mathbf{f}_R^{(\mathrm{j-1,j})} = \lambda_x\mathbf{f}_R^{(\mathrm{j-1,j})}.
\end{equation}
or alternatively:
\begin{equation}
    \mathbf{f}_R^{(\mathrm{j-1,j})} = e^{-\mathrm{i}\mu_x} \mathbf{f}_R^{(\mathrm{j,j})} = \lambda_x^{-1}\mathbf{f}_R^{(\mathrm{j,j})}.
\end{equation}
Substitution of this last equation into Eq.~\ref{eq:force_app} gives the resulting force balance as given in equation~\textcircled{1} of Eqs.~(\ref{eq:force_bal}) in which the superscripts are suppressed.\
Equations~\textcircled{2} and \textcircled{3} of Eqs.~(\ref{eq:force_bal}) can be derived similarly by expressing the force balance of, respectively, the bottom boundary and the bottom-left boundary and expressing these in terms of the forces of the $(\mathrm{j,j})^{\mathrm{th}}$ cell.

\section{Matlab code}
\label{app:matlab}

\begin{lstlisting}[columns=flexible , style=Matlab-editor, numbers=left, basicstyle=\tiny]
%%%%%%%%%%%%%%%%%%%%%%%%%%%%%%%%%%%%%%%%%%%%%%%
%%%% INVERSE APPROACH FOR DISPERSION CURVE CALCULATIONS %%%%
%%%%%%%%%%%%%%%%%%%%%%%%%%%%%%%%%%%%%%%%%%%%%%%
%% INPUT DATA (definition of the problem)
% UC size
Lx = 0.05;                           % UC size x-direction [m]
Ly = 0.05;                           % UC size y-direction [m]
Lz = 0.005;                          % UC size z-direction [m]
% Material
E = 210e9;                           % Young's modulus [Pa]
v = 0.3;                             % Poisson ratio [-]
rho = 7800;                          % Mass density [kg/m^3]
% Mesh definition
n_elx = 10;                          % Number of elements in x-dir [-] (>0)
n_ely = 10;                          % Number of elements in y-dir [-] (>0)
n_elz = 3;                           % Number of elements in z-dir [-] (>0)
% Addition to UC bare plate
scatterer = 'none';                  % 'none', 'resonator', 'mass'
f_res = 2500;                        % resonance of resonator [Hz]
m_ratio = 0.3;                       % Scatterer mass ratio vs plate [-]
% Parameters for dispersion curves
cont_name = {'O','A','B','O'};       % Name of the IBC contour
cont_co =  [0,0; pi,0; pi,pi; 0,0];  % IBC definition
step_size = 0.01*pi;                 % Resolution propagation constant
n_curves = 10;                       % Number of calculated wave modes 

%% FE preprocessing
% Element size
ax = Lx/n_elx;
ay = Ly/n_ely;
az = Lz/n_elz;
% Information UC
n_elem = n_elx*n_ely*n_elz;   
n_nodes = (n_elx+1)*(n_ely+1)*(n_elz+1);
nDOF = 3;                           % Number of DOFs per node;
n_DOFs = nDOF*n_nodes;
mass_UC = rho*(n_elx*ax)*(n_ely*ay)*(n_elz*az);
% Element matrices
[KE,ME] = KM(E,v,rho,ax,ay,az);
% FEM information (hard-coded for nDOF = 3)
nodeNrs = reshape(1:n_nodes, 1+n_ely, 1+n_elz, 1+n_elx); % nodes numbering
cMat = reshape(nDOF * nodeNrs(1:n_ely, 1:n_elz, 1:n_elx)+1, n_elem, 1) + ...
    [0,1,2,3*(n_ely+1)*(n_elz+1)+[0,1,2,-3,-2,-1],-3,-2,-1,3*(n_ely + ...
    1)+[0,1,2],3*(n_ely+1)*(n_elz+2)+[0,1,2,-3,-2,-1],3*(n_ely+1)+[-3,-2,-1]]; % connectivity matrix DOFs

%% DOFs partitioning
DofNrs = reshape(1:n_DOFs, nDOF*(n_ely+1), n_elz+1, n_elx+1);
dofs.L  = DofNrs(nDOF+1:end-nDOF,:,1); dofs.L = dofs.L(:);
dofs.R  = DofNrs(nDOF+1:end-nDOF,:,end); dofs.R = dofs.R(:);
dofs.T  = DofNrs(1:nDOF,:,2:end-1); dofs.T = dofs.T(:);
dofs.B  = DofNrs(end-(nDOF-1):end,:,2:end-1); dofs.B = dofs.B(:);
dofs.TL = DofNrs(1:nDOF,:,1); dofs.TL = dofs.TL(:);
dofs.TR = DofNrs(1:nDOF,:,end); dofs.TR = dofs.TR(:);
dofs.BL = DofNrs(end-(nDOF-1):end,:,1); dofs.BL = dofs.BL(:);
dofs.BR = DofNrs(end-(nDOF-1):end,:,end); dofs.BR = dofs.BR(:);
dofs.I  = DofNrs(nDOF+1:end-nDOF,:,2:end-1); dofs.I = dofs.I(:);

%% System matrix assembly
% Assemble bare plate system matrices
iL = reshape(kron(cMat,ones(24,1))',24*24*n_elem,1);
jL = reshape(kron(cMat,ones(1,24))',24*24*n_elem,1);
sK = reshape(KE(:)*(ones(n_elem,1)'),24*24*n_elem,1);
sM = reshape(ME(:)*(ones(n_elem,1)'),24*24*n_elem,1);
K = sparse(iL,jL,sK); 
M = sparse(iL,jL,sM); 
% Add scatterer (point mass/resonator)
switch scatterer
    case 'mass' 
      mass = m_ratio*mass_UC; 
      dofM = 3*nodeNrs(ceil(end/2),end,ceil(end/2));
      M(dofM,dofM) = M(dofM,dofM) + mass;
    case 'resonator'
      % Add extra zero row and column to K, M 
      K = [K zeros(n_DOFs,1);zeros(1,n_DOFs) 0];
      M = [M zeros(n_DOFs,1);zeros(1,n_DOFs) 0];
      % Define mass and stiffness
      mass = m_ratio*mass_UC;
      k_res = (f_res*2*pi)^2*mass; 
      dofK = 3*nodeNrs(ceil(end/2),end,ceil(end/2));
      n_DOFs = n_DOFs+1;
      dofM = n_DOFs;
      % Add resonator matrices to bare plate UC system matrices
      K([dofK dofM],[dofK dofM]) = K([dofK dofM],[dofK dofM]) + [k_res -k_res; -k_res  k_res];
      M([dofK dofM],[dofK dofM]) = M([dofK dofM],[dofK dofM]) + [0 0; 0 mass];
      dofs.I = [dofs.I; dofM];
end

%% Sampling the IBC
mu = 1i*cont_co(1,:).';
tot_steps = zeros(1,size(cont_co,1));
for i = 1 : size(cont_co,1)-1
    n_steps = ceil((sqrt((cont_co(i,1)-cont_co(i+1,1))^2+(cont_co(i,2)-cont_co(i+1,2))^2))/(step_size));
    ed = zeros(2,n_steps);
    for j = 1:2 % j = 1 -> x; j = 2 -> y
        step = (cont_co(i+1,j)-cont_co(i,j))/(n_steps); 
        if step == 0 
            ed(j,:) = cont_co(i,j)*ones(1,n_steps); 
        else 
            ed(j,:) = (cont_co(i,j)+step):step:cont_co(i+1,j); 
        end
    end
    mu = [mu, 1i*ed];
    tot_steps(i+1) = tot_steps(i)+n_steps;
end

%% Dispersion curve calculation
omega = zeros(n_curves,size(mu,2));
for i = 1:size(mu,2)
    % Construction of periodicity matrix
    R = eye(n_DOFs,n_DOFs); 
    R(dofs.R,dofs.L) = exp(mu(1,i))*eye(length(dofs.R),length(dofs.L));
    R(dofs.T,dofs.B) = exp(mu(2,i))*eye(length(dofs.T),length(dofs.B));
    R(dofs.TL,dofs.BL) = exp(mu(2,i))*eye(length(dofs.TL),length(dofs.BL));
    R(dofs.TR,dofs.BL) = exp(mu(1,i)+mu(2,i))*eye(length(dofs.TR),length(dofs.BL));
    R(dofs.BR,dofs.BL) = exp(mu(1,i))*eye(length(dofs.BR),length(dofs.BL));
    R = sparse(R(:,setdiff(1:n_DOFs,[dofs.R;dofs.T;dofs.TL;dofs.BR;dofs.TR])));
    % Impose periodicity boundary conditions
    K_BF = R'*K*R; 
    M_BF = R'*M*R; 
    % Compute dispersion curves
    [~,s] = eigs(K_BF,M_BF,n_curves,0);
    [s,~] = sort(diag(s));
    omega(:,i) = sqrt(s);
end

%% Plot
figure
plot(0:tot_steps(end),real(omega(:,:))/(2*pi),'ko','LineWidth',1);
xlabel('Re(\mu) [-]'); ylabel('Frequency [Hz]'); axis tight;
set(gca,'XTick',tot_steps(:),'XTickLabel',cont_name,'XGrid','on')

%% Function of element matrices
function [KE,ME] = KM(E,nu,rho,a,b,c)
syms x y z
C = (1/((1+nu)*(1-2*nu)))*[1-nu,nu,nu,0,0,0;
    nu,1-nu,nu,0,0,0;
    nu,nu,(1-nu),0,0,0;
    0,0,0,(1-2*nu)/2,0,0;
    0,0,0,0,(1-2*nu)/2,0;
    0,0,0,0,0,(1-2*nu)/2];
N1 = (1/(8*a*b*c))*(a-x)*(b-y)*(c-z);
N2 = (1/(8*a*b*c))*(a+x)*(b-y)*(c-z);
N3 = (1/(8*a*b*c))*(a+x)*(b+y)*(c-z);
N4 = (1/(8*a*b*c))*(a-x)*(b+y)*(c-z);
N5 = (1/(8*a*b*c))*(a-x)*(b-y)*(c+z);
N6 = (1/(8*a*b*c))*(a+x)*(b-y)*(c+z);
N7 = (1/(8*a*b*c))*(a+x)*(b+y)*(c+z);
N8 = (1/(8*a*b*c))*(a-x)*(b+y)*(c+z);

B = [diff(N1,x),0,0,diff(N2,x),0,0,diff(N3,x),0,0,diff(N4,x),0,0,diff(N5,x),0,0,diff(N6,x),0,0,diff(N7,x),0,0,diff(N8,x),0,0;
    0,diff(N1,y),0,0,diff(N2,y),0,0,diff(N3,y),0,0,diff(N4,y),0,0,diff(N5,y),0,0,diff(N6,y),0,0,diff(N7,y),0,0,diff(N8,y),0;
    0,0,diff(N1,z),0,0,diff(N2,z),0,0,diff(N3,z),0,0,diff(N4,z),0,0,diff(N5,z),0,0,diff(N6,z),0,0,diff(N7,z),0,0,diff(N8,z);
    diff(N1,y),diff(N1,x),0,diff(N2,y),diff(N2,x),0,diff(N3,y),diff(N3,x),0,diff(N4,y),diff(N4,x),0,diff(N5,y),diff(N5,x),0,diff(N6,y),diff(N6,x),0,diff(N7,y),diff(N7,x),0,diff(N8,y),diff(N8,x),0;
    0,diff(N1,z),diff(N1,y),0,diff(N2,z),diff(N2,y),0,diff(N3,z),diff(N3,y),0,diff(N4,z),diff(N4,y),0,diff(N5,z),diff(N5,y),0,diff(N6,z),diff(N6,y),0,diff(N7,z),diff(N7,y),0,diff(N8,z),diff(N8,y);
    diff(N1,z),0,diff(N1,x),diff(N2,z),0,diff(N2,x),diff(N3,z),0,diff(N3,x),diff(N4,z),0,diff(N4,x),diff(N5,z),0,diff(N5,x),diff(N6,z),0,diff(N6,x),diff(N7,z),0,diff(N7,x),diff(N8,z),0,diff(N8,x)];
KE = int(int(int(B.'*C*B, z,-c,c),y,-b,b),x,-a,a);
KE = E*double(KE);
[~, kua, kaa, kau] = FEM_incompatible_modes(E,nu,a,b,c);
KE = KE - kua*(kaa\kau);
KE = KE/2;

N = [N1,0,0,N2,0,0,N3,0,0,N4,0,0,N5,0,0,N6,0,0,N7,0,0,N8,0,0;
     0,N1,0,0,N2,0,0,N3,0,0,N4,0,0,N5,0,0,N6,0,0,N7,0,0,N8,0;
     0,0,N1,0,0,N2,0,0,N3,0,0,N4,0,0,N5,0,0,N6,0,0,N7,0,0,N8];
ME = int(int(int(N.'*N, z,-c,c),y,-b,b),x,-a,a);
ME = rho*double(ME/8);
end

%%%%%%%%%%%%%%%%%%%%%%%%%%%%%%%%%%%%%%%%%%%
%%% This Matlab code was written in October 2023                        
%%% by V. Cool, E. Deckers, L. Van Belle and C. Claeys                  
%%% Department of Mechanical Engineering, 3001 Heverlee, Belgium        
%%% Any comments are welcome: claus.claeys@kuleuven.be                  
%%%                                                                     
%%% The code is intended for educational purposes.                      
%%% Theoretical details are discussed in the corresponding paper:       
%%% "A guide to numerical dispersion curve calculations:                
%%% explanation, interpretation and basic Matlab code"                  
%%%                                                                     
%%% This code can be dowloaded from the corresponding Github page:      
%%% github.com/LMSD-KULeuven/2D_InverseUndamped_DispersionCurves             
%%%                                                                     
%%%                                                                     
%%% Copyright (c) 2023 KU Leuven Mecha(tro)nic System Dynamics (LMSD)        
%%%                                                                     
%%% Permission is hereby granted, free of charge, to any person         
%%% obtaining a copy of this software and associated documentation      
%%% files (the "Software"), to deal in the Software without             
%%% restriction, including without limitation the rights to use,        
%%% copy, modify, merge, publish, distribute, sublicense, and/or sell   
%%% copies of the Software, and to permit persons to whom the           
%%% Software is furnished to do so, subject to the following            
%%% conditions:                                                         
%%%                                                                     
%%% The above copyright notice and this permission notice shall be      
%%% included in all copies or substantial portions of the Software.     
%%%                                                                     
%%% THE SOFTWARE IS PROVIDED "AS IS", WITHOUT WARRANTY OF ANY KIND,     
%%% EXPRESS OR IMPLIED, INCLUDING BUT NOT LIMITED TO THE WARRANTIES     
%%% OF MERCHANTABILITY, FITNESS FOR A PARTICULAR PURPOSE AND            
%%% NONINFRINGEMENT. IN NO EVENT SHALL THE AUTHORS OR COPYRIGHT         
%%% HOLDERS BE LIABLE FOR ANY CLAIM, DAMAGES OR OTHER LIABILITY,        
%%% WHETHER IN AN ACTION OF CONTRACT, TORT OR OTHERWISE, ARISING        
%%% FROM, OUT OF OR IN CONNECTION WITH THE SOFTWARE OR THE USE OR       
%%% OTHER DEALINGS IN THE SOFTWARE.                                     
%%%%%%%%%%%%%%%%%%%%%%%%%%%%%%%%%%%%%%%%%%%%
\end{lstlisting}

\section{Details on the Matlab implementation}
\label{sec:matlab_impl}
In Sec.~\ref{sec:Matlab_impl_new}, an overview is given on how to use the \textsc{Matlab} implementation together with three initial case studies.\  
In what follows, the different sections of the \textsc{Matlab} code are discussed in more detail for the interested reader who wants to adapt or extend the code.\
The code is structured as follows; after the user-inputs, the FE model is constructed, the periodicity boundary conditions are applied and the dispersion curves are computed.\
More specifically, the code consists of the following building blocks:
\begin{enumerate}
    \item FE preprocessing (lines 27-44)
    \item DOFs partitioning (lines 46-56)
    \item System matrix assembly (lines 58-86)
    \item Sampling the IBC (lines 88-104)
    \item Dispersion curve calculation (lines 106-130)
\end{enumerate}
The different parts are discussed in the remainder of this appendix.\

\subsection{FE preprocessing (lines 27-44)}
\begin{figure*}
	\centering
	\includegraphics[width = 0.3\linewidth]{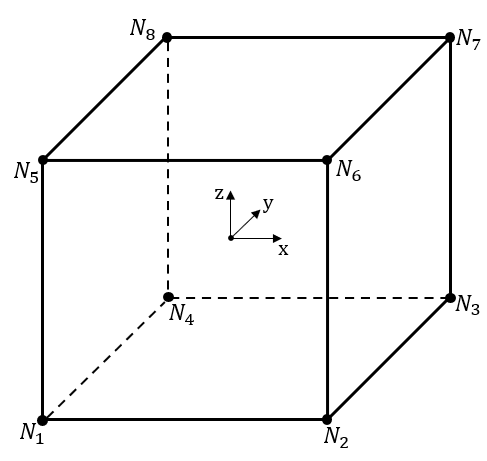}
	\caption{Numbering of the shape functions to derive the element stiffness and mass matrix.}
	\label{fig:el}
\end{figure*}

This section in the code prepares the FE discretization.\ 
First, a number of dependent variables are derived, namely the element sizes in $x$-, $y$- and $z$-direction, denoted by \verb+ax+, \verb+ay+ and \verb+az+, respectively.\
Also information of the UC FE model is derived: \verb+n_elem+ the number of elements in the mesh, \verb+n_nodes+ the number of nodes, \verb+n_DOFs+ the number of DOFs in the FE mesh and \verb+mass_UC+ the mass of the UC which is required if a scatterer is added to the plate.\
The next step is to compute the element stiffness (\verb+KE+) and mass (\verb+ME+) matrices, which is done by calling the function \verb+KM+ (defined on lines 132-167):
\begin{lstlisting}[columns=flexible, numbers=left, firstnumber=39, style=Matlab-editor, basicstyle=\small]
[KE,ME] = KM(E,v,rho,ax,ay,az)
\end{lstlisting}
Since a structured mesh is used with elements of equal size, only one element stiffness and mass matrix needs to be calculated.\
Generally, each element will have different element matrices.\
In the \verb+KM+ function, the constitutive matrix (\verb+C+) is first defined.\
Next, the shape function sequence is given as shown in Fig.~\ref{fig:el}.\ 
\verb+KE+ and \verb+ME+ are derived by full integration over the element with the symbolic \textsc{Matlab} toolbox{\footnote{For readers without access to the symbolic toolbox of \textsc{Matlab}, the element stiffness matrix on line~157 can be adapted to the first output kuu of the function FEM\_incompatible\_modes. The mass matrix should be implemented analogously.}} (lines 150-157 and 162-166).\
This could also be implemented with a numerical integration scheme using e.g.\ Gauss quadrature~\cite{zienkiewicz2005finite,cook2007concepts}.\
Care should be taken when linear solid elements are used to model thin plates if the elements' aspect ratio differs from 1, since a shear locking phenomenon can occur, resulting in a non-physical stiffness overestimation in bending.\
To deal with this shear locking, so-called incompatible mode elements are used in this work.\
The required changes to the element stiffness matrix are computed with the function \verb+FEM_incompatible_modes+ on line 158, inspired by the open access code available in~\cite{Bower}.\
For completeness, the function \verb+FEM_incompatible_modes+ is added to the \textit{Supplementary Material} of this paper.\
For more information regarding the derivation of the element matrices, the reader is referred to standard FE reference works, e.g.~\cite{cook2007concepts,zienkiewicz2005finite}.\

\begin{figure*}
	\centering
	\includegraphics[width = 0.9\linewidth]{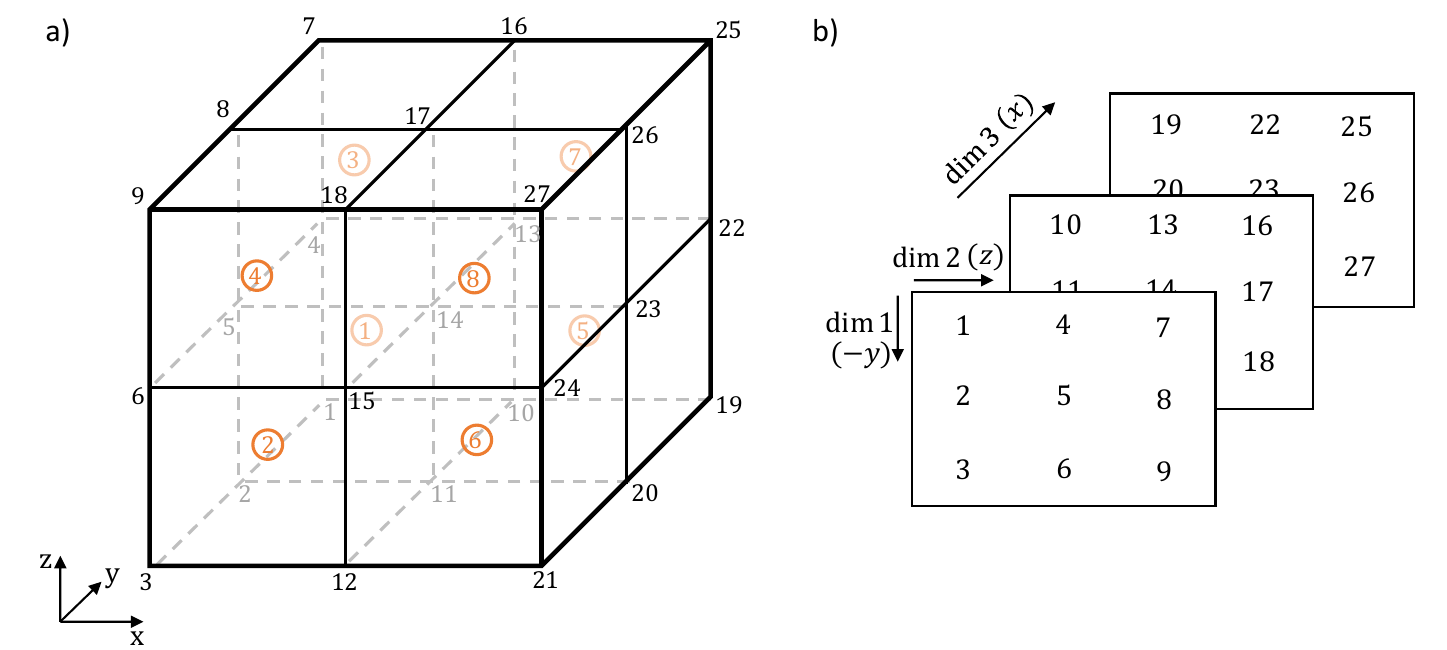}
	\caption{ a) Visualization of the elements and node numbering definition. b) Visualization of the variable $\mathrm{nodeNrs}$ including the node numbering.}
	\label{fig:FE1}
\end{figure*}

Next, the information required for assembling the FE model from the element matrices is derived.\ 
This code is based on the work of Ferrari et al.~\cite{ferrari2020new}.\
In Fig.~\ref{fig:FE1}a, the numbering of the elements (circled numbers) and nodes (black numbers) is given.\ 
The numbering increments according to the $-y$-, $z$- and then $x$-direction.\
The information of the node numbering is included in the \verb+nodeNrs+ tensor, which is visualized for a $2 \times 2 \times 2$ mesh in Fig.~\ref{fig:FE1}b.\
This is a three-dimensional tensor in which the first, second and third dimension follow the nodes in the $-y$-, $z$- and $x$-direction, respectively.\
Next, the connectivity matrix \verb+cMat+ is derived on lines 42-44 which contains for each element a row with the DOFs in the right order corresponding to the element matrix definition (cf.~Fig.~\ref{fig:el}).\ 
For the $2 \times 2 \times 2$ mesh in Fig.~\ref{fig:FE1}a, a snapshot of the \verb+cMat+ is given by:
\setcounter{MaxMatrixCols}{15}
\begin{equation}
    \mathrm{cMat} = \left[ \begin{matrix} 
       4  & 5 & 6 & 31 & 32 & 33 & \dots & 37 & 38 & 39 & 10 & 11 & 12 \\
       7  & 8 & 9 & 34 & 35 & 36 & \dots & 40 & 41 & 42 & 13 & 14 & 15 \\
          &   &   &   &   &   & \vdots &  &  &  &  &  &  \\
       40  & 41 & 42 & 67 & 68 & 69 & \dots & 73 & 74 & 75 & 46 & 47 & 48 \\
       43  & 44 & 45 & 70 & 71 & 72 & \dots & 76 & 77 & 78 & 49 & 50 & 51 \\
    \end{matrix} \right].
\end{equation}
For example, the first row is composed of the DOFs corresponding to the first element.\ 
As can be seen in Fig.~\ref{fig:FE1}, the nodes $1,2,4,5,10,11,13,14$ belong to this element.\
However, following the sequence of the shape functions (Fig.~\ref{fig:el}), this becomes $2,11,10,1,5,14,13,4$.\
Each node contains 3 DOFs, so node 1 corresponds to DOFs $1,2,3$, node $2$ to DOFs $4,5,6$ etc.\ 
cMat now contains these DOFs in the right order.\
Note, only the DOFs for nodes $2,11,13,4$ are visualized here in the first row of cMat to keep the matrix visualization digestible.\
With the \verb+cMat+ construction completed, the full stiffness and mass matrices of the system can be  assembled.

\subsection{DOFs partitioning (lines 46-56)}
\label{subsec:DOF_class}
\begin{figure*}
	\centering
	\includegraphics[width = 0.9\linewidth]{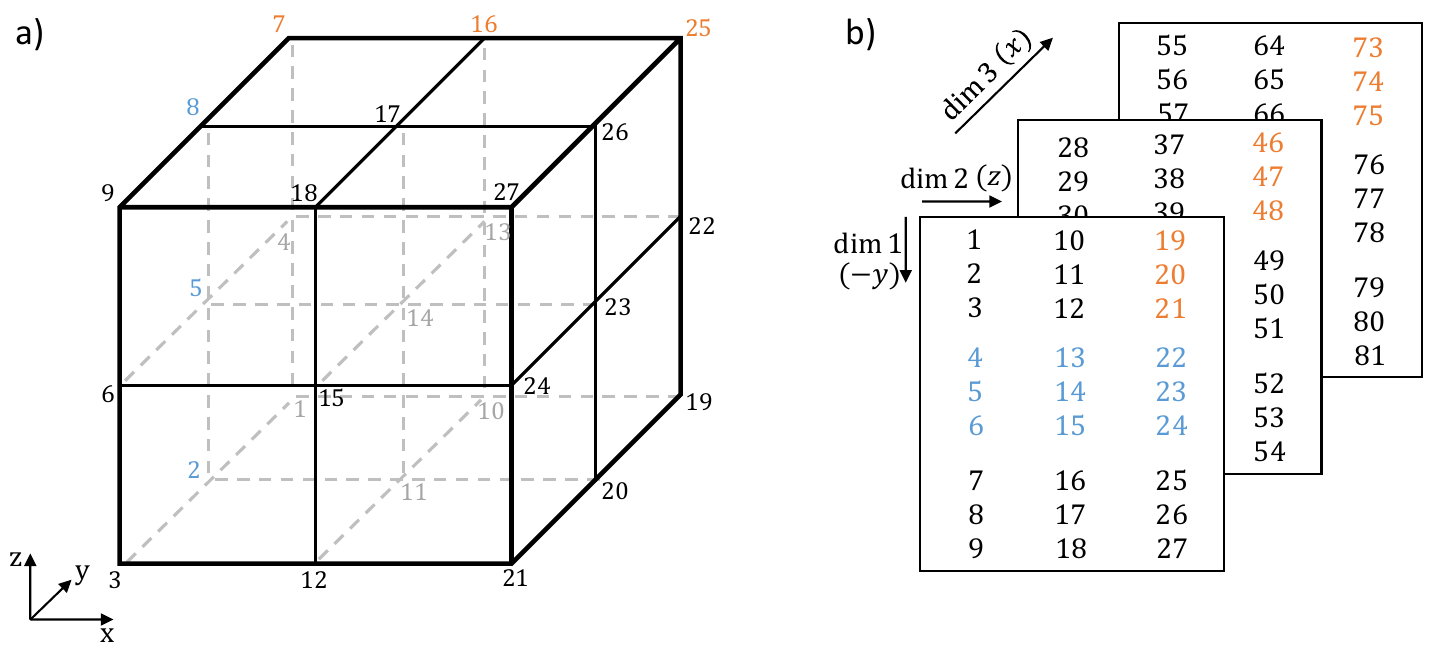}
	\caption{a) Node numbering visualization with two groups colored in blue and orange, b) visualization of the $\mathrm{DofNrs}$ tensor with the corresponding DOFs of the colored node groups.}
	\label{fig:FE2}
\end{figure*}

In order to apply the periodicity boundary conditions, a partitioning of the DOFs is required, cf.~Eq.~(\ref{eq:BF_BCs}).\
This is facilitated by defining the three-dimensional tensor \verb+DofNrs+ in a similar way as \verb+nodeNrs+.\
The \verb+DofNrs+ contains an ordering of the DOFs in the $-y$, $z$, $x$-direction respectively for dimension one, two and three, as illustrated in Fig.~\ref{fig:FE2}b for the $2 \times 2 \times 2$ example of Fig.~\ref{fig:FE2}a.\
This structure provides an easy way to select the DOFs which belong to certain faces or edges.\
This is visualized by the blue and orange node group in Fig.~\ref{fig:FE2}a which results in the blue and orange DOFs highlighted in Fig.~\ref{fig:FE2}b.\
More specifically, the blue nodes correspond to the middle $y$-node, first $x$-node and all $z$-nodes.\ 
This leads to the selection of the corresponding DOFs as rows $4-6$ in the first dimension, all $z$ columns in the second dimension and the first face in the third dimension, i.e.~\verb+DofNrs(4:6,:,1)+.\
Analogously, the DOF groups needed in Eq.~(\ref{eq:BF_BCs}) can be easily derived.\ 
Two examples are given here:
\begin{lstlisting}[columns=flexible, numbers=none, style=Matlab-editor, basicstyle=\small]
dofs.L  = DofNrs(nDOF+1:end-nDOF,:,1); dofs.L = dofs.L(:);
dofs.TR = DofNrs(1:nDOF,:,end); dofs.TR = dofs.TR(:);
\end{lstlisting}
in which the left DOFs \verb+dofs.L+ (colored blue in Fig.~\ref{fig:FE2}b) are defined as the DOFs corresponding to all nodes except the first and last one according to the $y$-direction, all nodes in the $z$-direction and the first node in the $x$-direction (colored blue in Fig.~\ref{fig:FE2}a).\
Correspondingly the top-right DOFs are defined as the DOFs corresponding to the first node in the -$y$-direction, all nodes in the $z$-direction and the last node in the $x$-direction.\

\subsection{System matrix assembly (lines 58-86)}
\label{subsec:sym_ass}
After the FE preprocessing, the required system matrices (\verb+K+,\verb+M+) of the UC FE model are assembled.\
First the system matrices of the plate are derived on lines 60-65, after which the information of the scatterer is added to the system matrices on lines 67-86 in case \verb+scatterer+ equals \verb+resonator+ or \verb+mass+.\
The system matrices of the plate are assembled using the \textsc{Matlab} function \verb+sparse+:
\begin{lstlisting}[columns=flexible, numbers=left, firstnumber=64, style=Matlab-editor, basicstyle=\small]
K = sparse(iL,jL,sK); 
M = sparse(iL,jL,sM); 
\end{lstlisting}
in which \verb+sK+ and \verb+sM+ contain all the coefficients of the element matrices over all elements in a column vector for the stiffness and mass matrix, respectively.\ 
\verb+iL+ and \verb+jL+ are a set of indices which map the information of \verb+sK+ and \verb+sM+ to the right position within the full system matrices \verb+K+ and \verb+M+, i.e.~$\mathrm{sK(i)}$ is located in the global matrix $\mathrm{K(iL(i),jL(i))}$~\cite{ferrari2020new}.\
These sets of indices are obtained using the information inside the \verb+cMat+ matrix.\

Next, the mass or resonator are added to the plate system matrices whenever the variable \verb+scatterer+ does not equal \verb+none+.\
For the point mass, this is done in lines~69-71: 
\begin{lstlisting}[columns=flexible, numbers=left, firstnumber=69, style=Matlab-editor, basicstyle=\small]
mass = m_ratio*mass_UC; 
dofM = 3*nodeNrs(ceil(end/2),end,ceil(end/2));
M(dofM,dofM) = M(dofM,dofM) + mass;
\end{lstlisting}
in which the first line determines the mass of the added point mass.\ 
The second line determines the DOF to which the point mass will be added, i.e.~here the $z$ DOF of the middle node at the top face of the plate.\ 
The corresponding node is easily determined with the \verb+nodeNrs+ structure, since now the node is selected in the middle of the $y$-axis and $x$-axis (obtained with \verb+ceil(end/2)+), while it is the final $z$-node (obtained with \verb+end+).\
The purpose of the 3 is to select the $z$-DOF of the corresponding selected node.\
The third line finally adds the selected mass to the right position in the mass matrix.\

For the resonator case, a mass-spring system is added to the UC, which corresponds to adding a single DOF system.\ 
First, the \verb+K+ and \verb+M+ matrices are extended with one zero row and one zero column to account for the added displacement DOF of the resonator mass.\
Next, the mass and stiffness of the mass-spring system are determined, whereby the mass is computed analogously to the point mass and the stiffness is computed using this mass and the selected frequency \verb+f_res+:
\begin{lstlisting}[columns=flexible, numbers=left, firstnumber=78, style=Matlab-editor, basicstyle=\small]
k_res = (f_res*2*pi)^2*mass; 
\end{lstlisting}
Similar to the point mass case, the mass-spring system is connected to the $z$-DOF of the middle node at the top face, denoted as \verb+dofK+.\ 
The resonator DOF is added as the last DOF of the extended \verb+K+ and \verb+M+ matrices, denoted by \verb+dofM+.
Finally, a $2 \times 2$ system of equations, representing the resonator and its interaction with the base structure, are added at the correct position within the total system matrices:
\begin{lstlisting}[columns=flexible, numbers=left, firstnumber=83, style=Matlab-editor, basicstyle=\small]
K([dofK dofM],[dofK dofM]) = K([dofK dofM],[dofK dofM]) + [k_res -k_res; -k_res  k_res];
M([dofK dofM],[dofK dofM]) = M([dofK dofM],[dofK dofM]) + [0 0; 0 mass];
\end{lstlisting}
Finally, on line 85, the new DOF (\verb+dofM+) is added to the interior part of the DOFs partitioning.\
If the reader would add a resonator at one of the boundaries of the UC, this line should be adapted to the corresponding DOF group.\

\subsection{Sampling the IBC (lines 88-104)}
\begin{figure*}
	\centering
	\includegraphics[width = 0.5\linewidth]{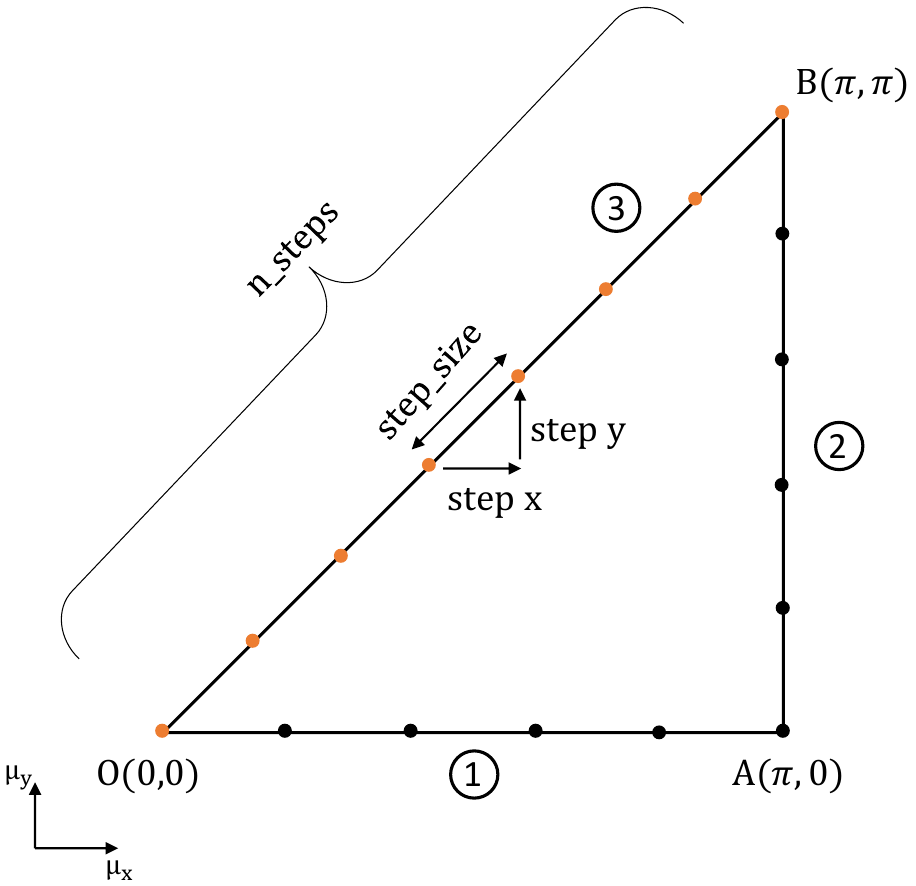}
	\caption{Visualization of how the IBC is sampled in the \textsc{Matlab} code.}
	\label{fig:IBC_code}
\end{figure*}

Before computing the dispersion curves, all $(\mu_x,\mu_y)$ pairs of interest are required along the IBC.\
These are sorted in the variable \verb+mu+.\ 
Fig.~\ref{fig:IBC_code} illustrates this for an example IBC of OABO.\
The code consists of a for-loop (lines 91-104) which loops over the different sections along the IBC.\ 
For the example of the OABO contour, these are the sections OA, AB and BO.\
This for-loop is preceded by initializing the \verb+mu+ vector with the first corner point, which corresponds to the first row of the \verb+cont_co+ variable.\
Also a vector \verb+tot_steps+ is initialized, which stores the number of steps for each section.\ 
This variable is only used for plotting.\
In the for-loop, the same computations are repeated for each section.
Consider for instance the section BO as an example (Fig.~\ref{fig:IBC_code}).\ 
First, the total number of steps in the section is computed, given by \verb+n_steps+:
\begin{lstlisting}[columns=flexible, numbers=left, firstnumber=92, style=Matlab-editor, basicstyle=\small]
n_steps = ceil((sqrt((contco(i,1)-contco(i+1,1))^2+(contco(i,2)-contco(i+1,2))^2))/(step_size));
\end{lstlisting}
The code computes the length of the section and divides it by the \verb+step_size+\ resolution.\ 
Next, the variable \verb+ed+ is computed which contains the $(\mu_x,\mu_y)$ pairs of the particular section, given by the orange dots in Fig.~\ref{fig:IBC_code}.\ 
This is done with another for-loop which looks at the $\mu_x$- and $\mu_y$ segmentation separately, respectively for \verb+j+ equal to one or two:
\begin{lstlisting}[columns=flexible, numbers=left, firstnumber=94, style=Matlab-editor, basicstyle=\small]
for j = 1:2 
    step = (contco(i+1,j)-contco(i,j))/(n_steps); 
    if step == 0 
        ed(j,:) = contco(i,j)*ones(1,n_steps); 
    else 
        ed(j,:) = (contco(i,j)+step):step:contco(i+1,j); 
    end
end
\end{lstlisting}
In this for-loop, first \verb+step+ is defined which is the current step-size taken in the $\mu_x$ or $\mu_y$ orthogonal coordinate space (see Fig.~\ref{fig:IBC_code}).\ 
Next, if this step is zero, which occurs in sections OA and AB, the current $\mu_x$ or $\mu_y$ does not change.\
If this step is not zero, the vector of $\mu_x$ or $\mu_y$ are computed on line 99.\
Finally, the computed \verb+ed+ is added to the \verb+mu+ variable and the \verb+tot_steps+ variable is supplemented.\
This is done for each section until the entire IBC is discretized.\
Now that all the to be evaluated $(\mu_x,\mu_y)$-pairs are defined, the dispersion curves can be computed.\

\subsection{Dispersion curve calculation (lines 106-130)}
Finally, the dispersion curves are computed along the IBC on lines 108-124.\
A for-loop is used to loop over the different ($\mu_x$,$\mu_y$)-pairs which are stored in the variable \verb+mu+.\
Note that the built-in parallel for-loop of \textsc{Matlab} (\verb+parfor+) can be used instead of a regular for-loop  to accelerate the computation.\
For each ($\mu_x$,$\mu_y$)-pair, the dispersion eigenvalue problem, cf.~Eq.~(\ref{eq:EVP_BF}), is solved to obtain the resulting frequencies.\

Solving the dispersion eigenvalue problem requires three steps:

(i)~On lines 110-116, the periodicity matrix \verb+R+, as defined in Eq.~(\ref{eq:BF_xy}), is computed.\
After initializing an identity matrix, it is filled with the periodicity boundary information, followed by the deletion of the DOFs which are not present in the periodic DOF vector (cf.~Eq.~(\ref{eq:BF_xy})).\

(ii)~The periodicity boundary conditions are imposed via a matrix-multiplication with the periodicity matrix \verb+R+ and \verb+R'+, cf.~Eq.~(\ref{eq:Proj}).\ 
This is implemented for the stiffness and mass matrices on lines 118 and 119, respectively:
\begin{lstlisting}[columns=flexible, numbers=left, firstnumber=118, style=Matlab-editor, basicstyle=\small]
 K_BF=R'*K*R; 
 M_BF=R'*M*R; 
\end{lstlisting}

(iii)~Finally the eigenvalue problem, cf.~Eq.~(\ref{eq:EVP_BF}), is solved to the eigenvalues with the built-in sparse \textsc{Matlab} eigenvalue solver \verb+eigs+:
\begin{lstlisting}[columns=flexible, numbers=left, firstnumber=121, style=Matlab-editor, basicstyle=\small]
[~,s] = eigs(K_BF,M_BF,n_curves,0);
\end{lstlisting}
in which \verb+n_curves+ defines the number of computed eigenvalues and the fourth input indicates the eigenvalues closest to zero are computed.\
Since the resulting eigenvalues in variable \verb+s+ correspond to $\omega^2$, an extra square-root operation allows obtaining the radial frequencies $\omega$, cf.~line~123.\

After the computations, the dispersion curves are visualized by plotting the frequencies against the evaluated points along the IBC (lines 127-130).\ 
A \verb+real+ operator is applied to the computed frequencies to delete possible numerical noise which can manifest as spurious imaginary parts of the frequencies.\ 
Via the division by $2\pi$, values in~[Hz] instead of [rad/s] are obtained.\
The x-axis labeling uses the variables \verb+tot_steps+ and \verb+cont_name+ to have a clear visualization of the used IBC.\

\section{Extensions}
\label{sec:ext}
For conciseness, the code as discussed above is limited to the inverse approach for the computation of dispersion diagrams of 2D periodic UCs with a possible addition of a point mass or resonator.\
Starting from this implementation, this appendix briefly elaborates on a number of possible extensions, which might be of interest to the reader.\ 
No detailed additional code is provided for all extensions, yet sufficient clues and references are given.\ 
Subsequently the following aspects are discussed:\\
\begin{enumerate}
    \item Wave mode calculation and plotting
    \item Adding multiple resonators per UC
    \item Computing the dispersion curves with the direct approach
    \item Adapting the script from 2D to 1D or 3D periodicity
    \item Computing a realizable resonator
    \item Accounting for non-orthogonal periodicity axes
    \item More comprehensive extensions such as model order reduction (MOR) or vibro-acoustic performance indicator calculations (i.e. sound transmission loss, vibration transmission etc.).
\end{enumerate}

\subsection{Wave mode calculation and plotting}
In the code, only the eigenvalues of the eigenvalue problem are computed, but not the eigenvectors which correspond to the wave modes.\
This can be easily extended to also compute and visualize the corresponding wave modes.\
The wave modes can also be calculated with the built-in function \verb+eigs+.\ 
To do so, lines 121 and 122 of the code are adapted to:
\begin{lstlisting}[columns=flexible, numbers=left, firstnumber=121, style=Matlab-editor, basicstyle=\small]
[Q,s] = eigs(K_BF,M_BF,n_curves,0);
[s,Idx] = sort(diag(s));
modeshapes{i} = Q(:,Idx);
\end{lstlisting}
The \verb+modeshapes+-variable now contains for each $(\mu_x,\mu_y)$-pair the displacement of all periodic DOFs $\tilde{\mathbf{q}}$ corresponding to the eigenvalues.\
Next, with Eq.~(\ref{eq:BF_xy}), the displacement of all DOFs of the UC can be extracted by expanding the wave modes over all UC DOFs.\

Plotting these wave modes in \textsc{Matlab} is straightforward.\ 
It requires the coordinates of the different nodes, which are easily deduced since a structured hexagonal mesh of equal-sized elements is used, together with the $x$-, $y$- and $z$-displacement of the nodes, which are incorporated in the displacement variable and ordered according to Fig.~\ref{fig:FE2}.\
The plotting in \textsc{Matlab} can be done with the function \verb+scatter+ or \verb+patch+.\
Note that the wave modes can also be plotted over several UCs since the displacement for the consecutive UCs is easily obtained by applying Bloch's theorem using the corresponding values for $(\mu_x,\mu_y)$, cf.~Eq.~(\ref{eq:qmu}).\

\subsection{Multiple resonators per UC}
In the literature, adding multiple resonators in one UC has been shown to enable several advantages, since the different resonators can target different frequencies leading to multiple bandgaps or diminish the characteristic performance decrease at the end of the stop band of a single resonator~\cite{claeys2016lightweight,claeys2017design,janssen2023improving}.\
When all resonators are represented by idealized mass-spring resonators, these cases can be easily implemented in the \textsc{Matlab} code.\
The main adaptation is how the stiffness matrix is adapted when the coefficients for the resonators are added, e.g.~lines 73-85.
First of all, the obtained \verb+Ks+ and \verb+Ms+ matrices should be enlarged with additional rows and columns corresponding to the amount of resonators, denoted with \verb+n_Res+, instead of only one:
\begin{lstlisting}[columns=flexible, numbers=left, firstnumber=74, style=Matlab-editor, basicstyle=\small]
 K = [K zeros(n_DOFs,n_Res); zeros(n_Res,n_DOFs+1)];
 M = [M zeros(n_DOFs,n_Res);zeros(n_Res,n_DOFs+1)]; 
\end{lstlisting}
Next, the mass, stiffness and attachment DOF for the spring should be defined for each resonator separately, after which their  corresponding $2 \times 2$ matrices should be added in the UC matrices at the right location.\
This could be implemented by replacing  lines 77-85 in the code with a for-loop over the different resonators:
\begin{lstlisting}[columns=flexible, numbers=left, firstnumber=77, style=Matlab-editor, basicstyle=\small]
for i = 1:n_Res
      % Define mass and stiffness
      mass(i) = ... 
      k_res(i) = ...
      dofK(i) = ...
      n_DOFs = n_DOFs+1;
      dofM = n_DOFs;
      % Add resonator matrices to UC mass, stiffness matrices
      K([dofK(i) dofM(i)],[dofK(i) dofM(i)]) = K([dofK(i) dofM(i)],[dofK(i) dofM(i)]) + [k_res(i) -k_res(i); -k_res(i)  k_res(i)];
      M([dofK(i) dofM(i)],[dofK(i) dofM(i)]) = M([dofK(i) dofM(i)],[dofK(i) dofM(i)]) + [0 0; 0 mass(i)];
      dofs.I = [dofs.I; dofM];
end
\end{lstlisting}
Note that when adding several resonators to the UC at separate locations, the symmetry of the UC can change and the IBC should be adapted accordingly on lines 22 and 23.\

\subsection{Direct calculation of dispersion curves}
Until now, the inverse approach was applied to compute the dispersion curves, resulting in freely propagating waves.\
Whenever the attenuation strength inside a bandgap is of interest, the direct approach $\boldsymbol\mu(\omega)$ is typically applied.\ 
This approach imposes real frequencies and solves the eigenvalue problem to the (complex) wave propagation constants.\
Keep in mind that the real and imaginary part of $\boldsymbol\mu$ correspond, respectively, with the amplitude and phase change when moving from one UC to the next.\
The methodology for computing dispersion curves with this approach is elaborated by Mace in~\cite{mace2005finite} and~\cite{mace2008modelling}, respectively, for $1$D periodic and $2$D periodic media.\
In this section, a general overview is given on how to adapt the \textsc{Matlab} code towards this approach.\

First of all, in the input section, the parameter of the dispersion diagram \verb+cont_co+ should be changed to the frequency range of interest to the user.\
The corresponding \verb+step_size+ is now instead defined as the frequency resolution.\
The code for the FE discretization and DOFs classification (lines 27-86) remains the same.\
The sampling of the IBC (lines 88-104) is not required anymore, since now the dispersion eigenvalue problem is solved for given frequencies.\
The dispersion curve calculation now requires a \verb+for+-loop over the desired frequencies, in which the corresponding eigenvalue problem is constructed for each frequency and subsequently solved to the wave propagation constant.\
More specifically, with the DOFs classification, sub-matrices can be extracted from the system matrices, e.g.
\begin{lstlisting}[columns=flexible, numbers=none, style=Matlab-editor, basicstyle=\small]
D = K - om^2 M;
D_II = D(dofs.I,dofs.I);
\end{lstlisting}
in which \verb+om+ is the imposed frequency, \verb+D+ is the dynamic stiffness matrix, and \verb+D_II+ is the submatrix of \verb+D+ which relates to the interior DOFs.\
The interior DOFs are often first eliminated using dynamic condensation before applying the periodic boundary conditions.\
More information on the construction of the eigenvalue problem can be found in \cite{mace2005finite} for 1D periodic media and \cite{mace2008modelling} for 2D periodic media.\ Note that, for 1D periodic media, after imposing the frequency, only one unknown remains and a quadratic eigenvalue problem needs to be solved.\
For 2D periodic media, after imposing the frequency, two unknown wave propagation constants remain.\
Therefore, different strategies have been presented which employ different formulations of the eigenvalue (e.g.\ polynomial or transcendental).\

After solving the eigenvalue problem, the results can be plotted.\
Again, different visualization strategies are encountered in the literature: (i)~a separate plot for the real and imaginary part of the wave numbers with respect to the frequency, e.g.~\cite{krushynska2016visco,hussein2014dynamics,mace2008modelling}, (ii) a 3D representation showing the imaginary part on the x-axis, real part on the y-axis and frequency on the z-axis, e.g.\ using \verb+scatter3+ in \textsc{Matlab}~\cite{krushynska2016visco,van2017impact,cool2022contribution}, (iii) a hybrid way, plotting the real part of the wave numbers while coloring the curves according to the imaginary part of the wave number~\cite{van2017impact,wang2015wave}.\

\subsection{1D, 3D periodicity}
While this manuscript explains the dispersion diagrams and provides code for structures which are periodic in two dimensions, i.e.~$x$- and $y$-direction, interesting UC designs have been proposed which are 1D or 3D periodic, e.g.~\cite{liu2000locally,delpero2016structural,sorokin2004plane,mace2005finite}.\
For such cases, the code can be adapted as follows:

(i)~In the input, mainly the definition of the IBC should be adapted on lines 22 and 23.\ 
The \verb+cont_co+ matrix should contain the same number of columns as the dimensionality of the periodicity.\ 
Based on the symmetry of the UC under investigation, the correct IBC should also be selected~\cite{kittel2005introduction,maurin2018probability}.\ 
    
(ii) The preparation of the FE discretization (lines 27-44) and the assembly of the system matrices (lines 58-65) can remain the same if simple rectangular structures are investigated, meshed with identical elements.\ 
Otherwise, for more general and geometrically complex structures, this part should be adapted as described in the next section.\
Note that the shear locking adaptations on line 158-159 can be deleted when this is not required.\

(iii) While adding the point mass or the mass-spring system, the DOF to which the mass or spring is added can also be adapted.\ 
Now this is encoded on line 70 and line 79 as:
         \begin{lstlisting}[columns=flexible, numbers=none, style=Matlab-editor, basicstyle=\small]
          3*nodeNrs(ceil(end/2),end,ceil(end/2))
         \end{lstlisting}
which means that the mass or spring is connected to the $z$-displacement DOF at the node in the middle of the top surface.\

\begin{figure*}
	\centering
	\includegraphics[width = \linewidth]{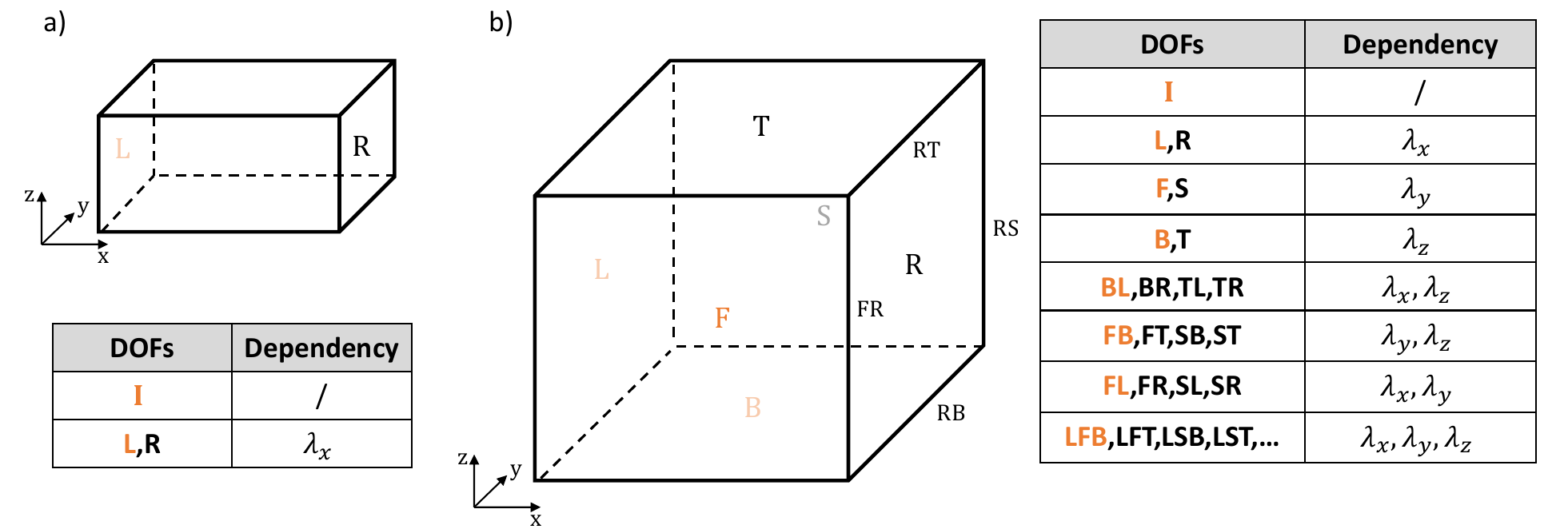}
	\caption{Schematic of the DOF groups and their dependency for the periodic boundary conditions for a) a 1D periodic UC case and b) a 3D periodic UC case.}
	\label{fig:3D_ext}
\end{figure*}

(iv)~The structuring of the DOFs should be adapted on lines 47-56.\ 
The different DOF groups for the 1D and 3D periodicity are visualized in Fig.~\ref{fig:3D_ext}.\ 
For the 1D periodicity this means that lines 50-55 can be deleted while the interior DOFs are now defined as:
\begin{lstlisting}[columns=flexible, numbers=none, style=Matlab-editor, basicstyle=\small]
dofs.I  = DofNrs(:,:,2:end-1); dofs.I = dofs.I(:);
\end{lstlisting}
For the 3D periodicity, extra relationships between the front and back, between the different edges and between the corner points should be defined, for which the same reasoning as in~\ref{subsec:DOF_class} can be followed.

(v)~During the sampling of the IBC (lines 88-104), most of the code can remain the same, apart from the considered dimensionality.\  
On line 92, the difference of \verb+cont_co+ between the different directions of periodicity should be taken, i.e.~one for the 1D and three for the 3D case.\
The inner loop should loop across all periodicity directions: \verb+j+ equals $1$ in the 1D periodic case and should loop from $1$ till $3$ for the 3D periodic case.\

(vi)~Finally, during the computation of the dispersion curves, the construction of the \verb+R+ matrix (lines 111-115) should be adapted to include all the required periodicity boundary conditions as visualized in Fig.~\ref{fig:3D_ext}.\
The transformation (lines 118-119) and the computation itself (lines 121-123) remain the same.

\subsection{Realizable resonator}
\begin{figure*}
	\centering
	\includegraphics[width = \linewidth]{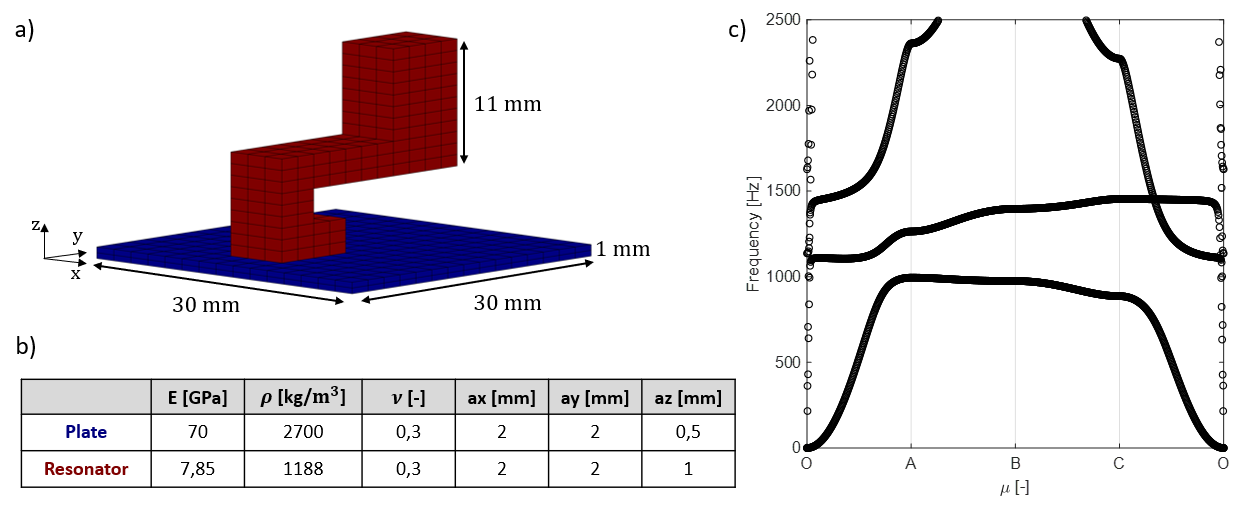}
	\caption{Example of a possible realizable resonator which can be implemented. a) The resonator design in which the different colors indicate the difference in element size and material, listed in the table b). c) The corresponding dispersion diagram. }
	\label{fig:Res_ext}
\end{figure*}

Until now, only plates with or without an additional point-mass or ideal mass-spring resonator are included.\
However, it could be of interest to investigate more complex resonator geometries.\
An example of a more complex UC design is given in Fig.~(\ref{fig:Res_ext}), together with the resulting dispersion curves.\
This UC is inspired on the work of Van Belle et al.~\cite{van2017impact}.\
Due to the generality of the code, this UC (and others) can be implemented while only changing slightly the FE part of the code (lines 27-44 and 58-86).\
Note, however, that for general complex UC designs, an unstructured body-fitted FE mesh can be of interest, for which the code has to be extended more extensively.\
All parts corresponding to the FE discretization will need to be extended (lines 27-44, 58-86 and 132-167), while the DOFs partitioning needs to be adapted correspondingly.\

In the example of the resonator in Fig.~\ref{fig:Res_ext}a, the structured hexagonal elements can be used with only a small extension to the code.\  
The plate, shown in blue, and the resonator, shown in red, consist of a different material and different element size, as given in Fig.~\ref{fig:Res_ext}).\
In the FE analysis this would result in defining two element sizes by adapting lines 29-31, together with obtaining the corresponding element matrices by adapting line 39.\
Next, the assembly of the full matrices will be more complicated since it needs to take into account the correct placement of the different element matrices which are no longer equal for all elements.\
Moreover, the \verb+ones(n_elem,1)+ command on lines 60 and 61 is not valid anymore since not all possible element positions in the rectangular $L_x \times L_y \times L_z$ domain are filled with material.\
After the correct \verb+K+ and \verb+M+ matrices of the UC are obtained, the DOFs structuring, sampling of the IBC and computation of the dispersion curves all remain the same as in the provided \textsc{Matlab} code.\

\subsection{Non-orthogonal periodicity axes}
In this paper only the case of orthogonal periodicity directions is considered.\

In the most general case, a UC is a parallelogon which can take 5 different shapes~\cite{kittel2005introduction} namely oblique, rectangular, rhombic, square, and hexagonal, with corresponding periodicity directions for each case~\cite{maurin2018probability}.\

For non-orthogonal periodicity directions, the FBZ will become hexagonal shaped and the basis vectors of the direct space ($\mathbf{d_i}$) and the basis vectors of the reciprocal wave space ($\mathbf{e_i}$)  are related by $\mathbf{d_i}\mathbf{e_j}=\delta_{ij}$, with $\delta_{ij}$ the Kronecker delta function and subscripts $i$ and $j$ taking integer values 1 and 2 for 2D structures~\cite{brillouin1946wave}.\
Although this influences somewhat the interpretation of dispersion diagrams, especially since the base vectors $\mathbf{e_1}$ and $\mathbf{e_2}$ in the reciprocal wave space are no longer orthogonal, the reasoning of this paper still applies.\ 

Regarding the \textsc{Matlab} code, the same structure and reasoning can be followed to evaluate the dispersion curves.\  
However, the required adaptations and extensions are more extensive since the simple, cuboid FE discretization cannot be applied anymore.\
The main parts which should be adapted are lines 27-44, 58-86 and 132-167.\
This can be implemented more generally with the help of standard FE handbooks~\cite{cook2007concepts,zienkiewicz2005finite} or reading in the FE meshes using a commercial preprocessor.\
Correspondingly, the DOFs partitioning (lines 46-56) should be adapted to the FE mesh.\
As before, also the coordinates of the IBC will be dependent on the symmetry of the UC~\cite{maurin2018probability}.\

\subsection{Elaborate extensions with the code as a starting point}
In this section, a few more elaborate extensions are briefly described and appropriate references are listed.\
Firstly, the computation of the dispersion curves can be computationally demanding, which the reader might discover when experimenting with the code.\
To alleviate this computational burden, model order reduction (MOR) techniques can be applied.\
The most commonly used techniques to reduce the computational cost for dispersion curve calculations are the Bloch mode synthesis (BMS)~\cite{krattiger2014bloch} and generalized BMS (GBMS)~\cite{krattiger2018generalized}.\
Both are model reduction techniques of the Craig-Bampton type, using modal information of the full order UC model, in order to represent it with a less expensive yet approximate reduced model.\
The BMS technique only reduces the interior UC DOFs, while the GBMS also reduces the boundary UC DOFs.\
The paper of Krattiger et al.~\cite{krattiger2014bloch} can be followed for the implementation of the BMS and \cite{krattiger2018generalized} for the GBMS.\
A brief description on how to the adapt the code is given here for the BMS.\ 
Up until the sampling of the IBC (lines 1-104), the basic \textsc{Matlab} implementation remains the same.\ 
After this point, the UC DOFs need to be partitioned in to an interior and boundary DOF subset, e.g. by constructing corresponding DOF groups \verb+dofs.I+ and \verb+dofs.A+, respectively.\
Next, the reduction basis is constructed following~\cite{krattiger2014bloch} as:
\begin{lstlisting}[columns=flexible, numbers=none, style=Matlab-editor, basicstyle=\small]
[Phi,~] = eigs(K(dofs.I,dofs.I),M(dofs.I,dofs.I),nRI,0);
Psi = -K(dofs.I,dofs.I)\K(dofs.I,dofs.A);
B = [Phi Psi; zeros(nA,nRI) eye(nA,nA];
\end{lstlisting}
with \verb+nRI+ the number of interior modal DOFs after the reduction, \verb+nA+ the (unreduced) number of boundary DOFs, \verb+Phi+ the interior normal modes and \verb+Psi+ the static constraint boundary modes.\
Afterwards, the reduced system matrices of the UC are obtained with the appropriate multiplication of the reduction basis \verb+B+:
\begin{lstlisting}[columns=flexible, numbers=none, style=Matlab-editor, basicstyle=\small]
M_red = B.'*M*B;
K_red = B.'*K*B;
\end{lstlisting}
These reduced UC matrices now should be used when applying the periodicity boundary conditions on line 118 and 119.\
Note that the construction of the matrix \verb+R+ needs to be adapted to the right number and ordering of the DOFs after the reduction.\ 
It is noted that for BMS, the amount of boundary DOFs remains te same. For the GBMS, this is not the case and additional care should be taken in renumbering and reordering the DOFs as well as in constructing the matrix \verb+R+.

Next, with the code as a starting point, many FE UC modeling based techniques can be implemented which compute performance indicators other than the dispersion diagrams.\ 
Important is that these techniques employ the wave and finite element (WFE) method which uses the UC of a periodic structure as a reference point and combines the FE technique with the infinite periodic theory.\
Following calculations are amongst others possible: (i) computation of the forced response of 2D homogeneous media, as studied by Renno et al.~\cite{renno2011calculating}, (ii) finite structure forced response computation of periodic media as studied by~\cite{van2022fast}, (iii) infinite periodic structure sound transmission loss (STL) e.g.~\cite{deckers2018prediction,parrinello2016transfer,xiao2021sound} etc., (iv)~finite structure STL, as studied by Yang et al.~\cite{yang2021wave} and (v)~wave mode contributions linking dispersion curves with the infinite periodic structure STL, as studied by Cool et al.~\cite{cool2022contribution}.\
Note that this list is not exhaustive and the reader is referred to the literature to find out more about these and other WFE -based techniques.\

\bibliography{mybibfile}

\end{document}